\documentclass[apj]{emulateapj}
\usepackage{graphicx}

\newcommand{\hi}          {\mbox{\rm \ion{H}{1}}}

\newcommand{\htwo}        {\mbox{H$_{2}$}}
\newcommand{\jone}        {$J=1\rightarrow0$}

\newcommand{\kmpers}      {\mbox{\rm km~s$^{-1}$}}
\newcommand{\percmcu}     {\mbox{\rm cm$^{-3}$}}
\newcommand{\msunperpcsq} {\mbox{\rm M$_\odot$~pc$^{-2}$}}
\newcommand{\xco}         {\mbox{$X_{\rm CO}$}}
\newcommand{\xcounits}    {\mbox{\rm cm$^{-2}$(K km s$^{-1}$)$^{-1}$}}
\newcommand{\Kkmpers}     {\mbox{\rm K km s$^{-1}$}}
\newcommand{\Kkmperspcsq} {\mbox{\rm K km s$^{-1}$ pc$^2$}}
\newcommand{\co}          {\mbox{$^{12}$CO}}
\newcommand{\Ico}         {\mbox{I$_{\rm CO}$}}
\newcommand{\av}          {\mbox{$A_V$}}
\newcommand{\percmsq}     {cm$^{-2}$}
\newcommand{\cii}         {\mbox{\rm [\ion{C}{2}]}}
\newcommand{\fscii}       {($^2$P$_{3/2}\rightarrow^2$P$_{1/2}$)}

\begin{document}

\title{The Resolved Properties of Extragalactic Giant Molecular Clouds}

\author{Alberto D. Bolatto\altaffilmark{1}, Adam K. Leroy\altaffilmark{2},
  Erik Rosolowsky\altaffilmark{3,4}, Fabian Walter\altaffilmark{2}, \& Leo
  Blitz\altaffilmark{5}}

\altaffiltext{1}{Department of Astronomy and Laboratory for
  Millimeter-wave Astronomy, University of Maryland, College Park, MD
  20742, USA}
\email{bolatto@astro.umd.edu}

\altaffiltext{2}{Max-Planck-Institut f\"ur Astronomie, D-69117
Heidelberg, Germany}

\altaffiltext{3}{Harvard-Smithsonian Center for Astrophysics, Cambridge, MA
  02138, USA}

\altaffiltext{4}{Department of Mathematics, Statistics, and Physics,
University of British Columbia at Okanagan, Kelowna, B.C. V1V 1V7,
Canada}

\altaffiltext{5}{Department of Astronomy and Radio Astronomy Laboratory,
  University of California at Berkeley, Berkeley, CA 94720,
  USA}

\begin{abstract}
We use high spatial resolution observations of CO to systematically
measure the resolved {\em size-line width}, {\em luminosity-line
width}, {\em luminosity-size}, and the {\em mass-luminosity} relations
of Giant Molecular Clouds (GMCs) in a variety of extragalactic
systems. Although the data are heterogeneous we analyze them in a
consistent manner to remove the biases introduced by limited
sensitivity and resolution, thus obtaining reliable sizes, velocity
dispersions, and luminosities. We compare the results obtained in
dwarf galaxies with those from the Local Group spiral galaxies. We
find that extragalactic GMC properties measured across a wide range of
environments are very much compatible with those in the Galaxy. The
property that shows the largest variability is their resolved
brightness temperature, although even that is similar to the average
Galactic value in most sources. We use these results to investigate
metallicity trends in the cloud average column density and virial
CO-to-H$_2$ factor.  We find that these measurements do not accord
with simple predictions from photoionization-regulated star formation
theory, although this could be due to the fact that we do not sample
small enough spatial scales or the full gravitational potential of the
molecular cloud. We also find that the virial CO-to-H$_2$ conversion
factor in CO-bright GMCs is very similar to Galactic, and that the
excursions do not show a measurable metallicity trend. We contrast
these results with estimates of molecular mass based on far-infrared
measurements obtained for the Small Magellanic Cloud, which
systematically yield larger masses, and interpret this discrepancy as
arising from large H$_2$ envelopes that surround the CO-bright
cores. We conclude that GMCs identified on the basis of their CO
emission are a unique class of object that exhibit a remarkably
uniform set of properties from galaxy to galaxy.
\end{abstract}


\section{Introduction}

Giant Molecular Clouds (GMCs) are the major reservoirs of molecular
gas and the sites of most star formation in our Galaxy and other
galaxies.  Their properties set the initial conditions for
protostellar collapse, and may play a role in determining the stellar
initial mass function \citep{MCKEE07}. Moreover, because GMCs provide
the bulk of the material for forming new stars their creation may be
the limiting mechanism that regulates the rate of star formation in
galaxies. Therefore, increasing our understanding of their properties and
distribution throughout the different environments of external
galaxies is likely to provide further insights into GMC and stellar
formation processes. There is a limited amount of information,
however, that can be gained from studies that resolve the general
distribution of molecular gas but not the individual molecular clouds.
Resolving GMCs --- to measure their sizes, velocity dispersions, and
luminosities --- is a critical step in understanding the processes
that ultimately drive galaxy evolution.

Studies of resolved molecular clouds in the Milky Way find that GMCs
are in approximate virial equilibrium and obey scaling relations,
commonly known as Larson laws \citep{LARSON81}, that have their origin
in the character of the turbulence in the interstellar medium
\citep{ELMEGREEN04,MCKEE07}.  Large scale surveys of the Milky Way
show that GMCs follow uniform scaling relations
\citep{SOLOMON87,ELMEGREEN96}, with few differences present between
those located in the inner and the outer disk \citep*{HEYER01}.

\citet{LARSON81} established that velocity dispersion, size, and
luminosity are correlated in Milky Way GMCs.  Observations
indicate that GMC line widths increase as a power of their radius
\citep{SOLOMON87}, such that

\begin{equation}
\sigma_v\approx0.72\, R^{0.5}\ {\rm km~s^{-1}}~,
\label{size_linewidth_eq}
\end{equation}

\noindent where $\sigma_v$ is the one-dimensional velocity dispersion
of the GMC, and $R$ is its radius measured in parsecs.  Equation
\ref{size_linewidth_eq} is known as the {\em size-line width}
relation, and it has been shown to hold even within GMCs down to
very small scales \citep{HEYER04,ROSOLOWSKY08}.

The {\em size-line width} relation is an expression of the equilibrium
turbulence conditions in the molecular ISM.  Detailed modeling of GMC
line profiles shows that the emission is macroturbulent (i.e., the
scale size of the turbulence is larger than the photon mean free
path), corresponding to many optically thick clumps that have a
clump-to-clump velocity dispersion similar to the observed line width
\citep{WOLFIRE93}. It is known that the observed line widths in all
but a few very compact, quiescent clouds are too large to be thermal;
at the typical GMC temperatures of $15-25$~K the thermal CO velocity
dispersion would be $\lesssim0.1$ \kmpers. They are rather due to
supersonic turbulence within the clouds. The dominant sources of
turbulence in the molecular ISM remain somewhat controversial. For
example, it is not yet known whether turbulence in molecular clouds is
primarily internally or externally driven
\citep[e.g.,][]{MCKEE07}. Contributors must include star formation ---
in the form of jets and winds for low mass stars, and winds and
expanding H{\small II} regions for massive stars --- and star
destruction --- in the form of supernovae shocks and expanding
superbubbles. Global processes that couple the large reservoirs of
thermal or rotational energy to the turbulent cascade --- such as
spiral shocks and several types of instabilities that involve shear,
magnetic fields, and self gravity --- are important on the largest
spatial scales. The fact that turbulence is present on scales
$\gtrsim100$ pc \citep[e.g.,][]{BRUNT03,DIB05}, and that the
dissipation lifetime of the largest eddies is $\lesssim10^7$ yr
\citep*{FLECK81,STONE98}, point to constant (or at least frequent)
energy injection on large scales.  This in turn suggests that massive
stellar death and gobal mechanisms are predominantly responsible for
the bulk of turbulent energy injection in the Milky Way, although
their respective dominance is a matter of debate
\citep{MACLOW04,PIONTEK04,PIONTEK05}.

The last two Larson relations, the {\em luminosity-line width} and {\em
luminosity-size} relations, describe correlations between cloud
luminosity, $L_{\rm CO}$, and either velocity dispersion or size
(note, as is frequently pointed out, that only two of the three Larson
relations are independent). They are \citep{SOLOMON87}

\begin {equation}
L_{\rm CO}\approx130\, \sigma_v^{5}\ \Kkmperspcsq,
\label{lum_lw_eq}
\end{equation}

\noindent and 

\begin{equation}
L_{\rm CO}\approx25\, R^{2.5}\ \Kkmperspcsq,
\label{lum_size_eq}
\end{equation} 

\noindent with $\sigma_v$ and $R$ expressed in their corresponding
units of \kmpers\ and pc.

Thus, the observations show that the virial mass, $M_{vir}$, relates
to the CO luminosity (which is proportional to the inferred luminous
mass $M_{lum}$) in the following manner \citep{SOLOMON87},

\begin{equation}
M_{vir}\approx39\, L_{\rm CO}^{0.81}\ {\rm M}_\odot 
\label{lco_mvir_eq}
\end{equation}

\noindent where $L_{\rm CO}$ is expressed in \Kkmperspcsq. We will
call this equation the Galactic {\em mass-luminosity} relation. Note
that the exponent of $L_{\rm CO}$ is somewhat uncertain, and we take
here the value reported by \citeauthor{SOLOMON87}. Although the
relation may not be precisely linear it is almost so, and simple
models of molecular clouds can explain the observed linearity
\citep*{DICKMAN86}. This approximate proportionality between $L_{\rm
CO}$ and $M_{vir}$ within the Galaxy allows the definition of an
empirical CO-to-H$_2$ conversion factor, \xco, that relates the
intensity of a GMC in the $^{12}$CO \jone\ transition to its H$_2$
column density ($N(\htwo)=\xco\Ico$) and ultimately its mass. For GMCs
around a median mass of $5\times10^5$ M$_\odot$ \citeauthor{SOLOMON87}
obtained a CO-to-H$_2$ factor that, when adjusted to the currently
accepted distance of 8.5~kpc to the Galactic Center, yields
$\xco\approx1.9\times10^{20}$ \xcounits\ \citep{MCKEE07}.  In the
Milky Way, this ``virial mass'' approach yields conversion factors
similar to those obtained from non-dynamical measurements, strongly
suggesting that clouds are in a dynamical state intermediate between
marginal gravitational binding and virial equilibrium \citep*[e.g.,
$\gamma$-ray results and modeling of the Galactic plane dust
continuum;][]{BLOEMEN84,STRONG96,DAME01}, with magnetic energy
introducing a small but non-negligible correction on the small scales
\citep[$\lesssim50\%$;][]{CRUTCHER99}.  Throughout this paper we adopt
$\xco=2\times10^{20}$ \xcounits\ as the Galactic conversion
factor. Note, however, that some nearby lines-of-sight at high
Galactic latitudes may exhibit significant larger values of this
coefficient \citep*[e.g.,][]{MAGNANI88,GRENIER05}.

Therefore, mass estimates of GMCs show that they are gravitationally
bound and in approximate virial equilibrium, which requires
$M_{vir}\sim R\,\sigma_v^2\propto M_{lum}$. As a consequence of
Equation \ref{size_linewidth_eq} and virial equilibrium, all GMCs have
approximately the same mean column density: $N_{\rm H}\approx
1.5\times10^{22}$ cm$^{-2}$, or an equivalent mass surface density
$\Sigma\approx170$ M$_\odot$~pc$^{-2}$ (after applying a 40\%
correction for the mass contribution of elements heavier than
Hydrogen). An equivalent result is that all Milky Way GMCs have
approximately the same mean visual extinction from edge to edge,
$A_V\sim7.5$, using the standard dust-to-gas ratio \citep*{BOHLIN78}.

\citet{MCKEE99} argues that Equation \ref{lco_mvir_eq} is a consequence
of the interplay between CO chemistry, the Milky Way radiation field,
the dust-to-gas ratio, and the local pressure of the ISM. As a result,
H$_2$ clouds observable in CO are at least marginally
bound. Furthermore, \citeauthor{MCKEE99} argues that the constancy of
$N_{\rm H}$ stems from the fact that clouds with lower column density
would not be bright in CO, while clouds with higher column density
would rapidly collapse.

Another consequence of the Larson relations is that Galactic GMCs have
low volume-averaged densities, $50<n({\rm H_2})<500$
cm$^{-3}$. Therefore their clumpiness must be high in order to excite
the rotational transitions of CO and of other commonly observed
high-density tracer molecules such as HCO$^+$, which require volume
densities in excess of 10$^4$ cm$^{-3}$. This clumpiness is probably
induced by supersonic turbulence
\citep*{PADOAN98,OSTRIKER99}. Further, the average density of a GMC
decreases with increasing mass, and as a result the free fall time in
the cloud increases. This led \citet{KRUMHOLZ05} to predict that
larger GMCs will form collapsing cores and stars less rapidly than
smaller ones.

The purpose of this study is to determine, {\em using a consistent
methodology}, the relations between size, velocity dispersion, and
luminosity for extragalactic GMCs that are encapsulated in the Milky
Way by the Larson laws.  We do so by studying GMCs in dwarf galaxies
and comparing them with those in spiral galaxies. Dwarf galaxies are
both numerous and nearby, and host physical conditions that fall
outside those explored by surveys of the Milky Way. The range of
parameter space thus probed provides an ideal laboratory to test
theories of cloud structure and formation. Particularly, dwarf
galaxies have low metallicities and their lack of internal extinction
results in intense radiation fields, reminiscent of the conditions in
primitive galaxies. They display irregular morphologies and are not
dominated by spiral structure. Dwarfs often have slowly rising, nearly
solid-body, rotation curves; this translates into small rotational
shear and perhaps changes the sources and scales of turbulence in
their ISM. The stellar potential wells of dwarf galaxies are weaker
than those in large spiral galaxies and the ambient ISM may be
correspondingly less dense. These environmental changes may influence
the equilibrium conditions of GMCs, and affect their Larson relations.

The integrated properties of dwarf galaxies already show clear
evidence that environment affects at least the CO content. Dwarf
irregular galaxies are fainter in CO than large spiral
galaxies. Galaxies with metallicity $12 + \log {\rm O/H} \lesssim 8.0$
are seldom detected in CO emission \citep{TAYLOR98}. Dwarf galaxies of
higher metallicity detected in CO, however, show similar CO luminosity
relative to either their stellar or atomic gas content as larger
galaxies \citep{LEROY05}. Here we investigate whether the resolved
properties of GMCs in these galaxies also reflect their changing
environments.

\subsection{Resolving extragalactic molecular clouds}
\label{introresol}

The study of resolved GMC properties in other galaxies remains
technically challenging.  A typical size for a Galactic GMC is $\sim
40$~pc, which at the $\sim4$ Mpc distance of the M~81 group
translates to an angular size of $\sim2\arcsec$. Achieving such a
resolution requires an antenna 250 meters in diameter operating at
$\lambda=2.6$~mm, the wavelength of the ground transition of
$^{12}$CO. This is clearly an impractical proposition, but aperture
synthesis techniques used by millimeter-wave interferometers routinely
attain this resolution. They do so, however, at the price of reduced
surface brightness sensitivity with respect to their filled aperture
equivalents, and studies such as the one presented here are ultimately
limited by sensitivity. Until the advent of the Atacama Large
Millimeter Array (ALMA) these observations will remain challenging.

Because of these difficulties, the resolved properties of GMCs in
galaxies other than the Milky Way remain a largely unexplored domain.
Because of its proximity, located at a distance where even a modest
telescope can attain a resolution of $\sim30-40$~pc, the first galaxy
where observations capable of clearly resolving individual GMCs were
possible was the Large Magellanic Cloud \citep{ISRAEL82}. The first
resolved study of a GMC beyond the immediate vicinity of the Milky Way
was performed in M~31 \citep*{VOGEL87}, while the first large scale
interferometric survey of GMCs in another galaxy was done on M~33
\citep{WILSON90}. The first comprehensive analyses in the Magellanic
Clouds were performed on the Small Magellanic Cloud
\citep*{RUBIO91,RUBIO93b}.

\citet{BLITZ07} recently presented a study of GMCs in the Local Group,
analyzing observations from CO surveys complete down to a known mass
limit over a significant fraction of each galaxy. This study concluded
that the Larson relations in the Milky Way disk were approximately
those observed throughout the Local Group, although some offsets
(possibly methodological) were present. They also concluded that GMCs
within a particular galaxy have roughly constant surface density,
$\Sigma_{\rm GMC}$, and that once corrections for the local CO-to-H$_2$
conversion factor are applied the sample as a whole has a scatter of
only a factor of 2 in $\Sigma_{\rm GMC}$. Finally, they calculated a typical
conversion factor $\xco\approx4\times10^{20}$ \xcounits, accurate
to within 50\% for most of the sample except the Small Magellanic
Cloud, for which they obtain a value $\sim3$ times larger.

This paper proceeds as follows: in \S\ref{data} we describe the data
sets and and galaxies studied. In \S\ref{method} we summarize the
method used to consistently derive GMC properties across the different
datasets.  In \S\ref{results} we discuss the {\em size-line width},
{\em luminosity-line width}, and {\em luminosity-size} relations in
our sample of extragalactic GMCs, and compare them with the Galaxy. In
\S\ref{discussion} we discuss the implications of these results on the
equilibrium of clouds in the Small Magellanic Cloud, the
photoionization-regulated star formation theory, the CO-to-H$_2$
conversion factor, and the brightness temperature of extragalactic
clouds. In each case we contrast our results in dwarf galaxies with
those obtained in the Milky Way and the other Local Group spirals,
M~31 and M~33.  In \S\ref{summary} we summarize this study and present
our conclusions.

\section{Data}
\label{data}

We use data from four telescopes: the Berkeley-Illinois-Maryland Array
\citep[BIMA;][]{WELCH96}, Caltech's Owens Valley Radio Observatory
(OVRO) millimeter array, the Institut de Radioastronomie
Millim\'etrique Plateau de Bure Interferometer (PdBI), and the
Swedish-ESO Submillimeter Telescope (SEST). We measure GMC properties
in $11$ dwarf galaxies: the Large and Small Magellanic Clouds, IC~10,
NGC~185, NGC~205, NGC~1569, NGC~2976, NGC~3077, NGC~4214, NGC~4449,
and NGC~4605. We compare GMC properties derived in dwarfs to those
found in the Milky Way \citep[data obtained by the Five Colleges Radio
Astronomy Observatory, FCRAO;][]{SANDERS86} and the other two Local
Group spiral galaxies, M~31 and M~33. The latter two act as key
``control'' data sets because they are obtained by the same telescopes
at similar angular and spatial resolutions to the dwarf galaxy GMC
data.  Table \ref{DATATABLE} gives the source, telescope, line
observed, resolutions, and reference for each data set.

When we analyze observations of the CO $J=2\rightarrow1$ transition,
we assume $I (2\rightarrow1) = I (1\rightarrow0)$ (i.e., thermalized
optically thick emission). For the GMC complex in the N83 region of
the Small Magellanic Cloud, this is within $10\%$ of the observed
ratio \citep{BOLATTO03}. In the Milky Way the observed ratio varies
from unity to $I (2\rightarrow1) \sim~0.65 I (1\rightarrow0)$
\citep[e.g.,][]{SAWADA01}.

Several data sets appear here for the first time. Notably, GMCs in
NGC~3077, NGC~4214, and NGC~4449 where mapped at BIMA as part of a
large project studying the molecular ISM in dwarf galaxies. The BIMA
maps of NGC~2976 and NGC~4605 were obtained as part of the same
project and have appeared previously in kinematic analyses
\citep{BOLATTO02,SIMON03}. The observation and reduction procedure for
these data follow the descriptions in those papers.

Table \ref{GALAXYTABLE} lists the properties of the galaxies that we
study.  Most have late type morphologies; the exceptions are NGC~185
and NGC~205, Local Group dwarf ellipticals mapped by \citet{YOUNG01}
using BIMA. All of our sources are $<5$~Mpc away; the most distant
galaxy is NGC~4605 at 4.26~Mpc. We indicate the absolute blue
magnitude of each galaxy corrected by Galactic extinction to
illustrate their luminosity. The galaxies span the range $-18\lesssim
M_{B} \lesssim -15$~magnitudes. We give metallicities for each galaxy,
with the corresponding references for metallicity and distance.

\begin{center}
\begin{deluxetable}{lccccc}
\tabletypesize{\scriptsize}
\tablewidth{0pt}
\tablecolumns{4} 
\tablecaption{\label{DATATABLE} Data Sets
    Used In This Paper} 
\tablehead{ \colhead{Source} & \colhead{Instr.} &
    \colhead{Transition} & \colhead{Resol.} & \colhead{Resol.} & \colhead{Ref.} \\
& & & \colhead{(pc)} & \colhead{(\kmpers)} }
\startdata
  IC~10 & OVRO & $1-0$ & 19 & 0.7 & 5 \\
  NGC~185 & BIMA & $1-0$ & 15 & 4 & 1 \\
  NGC~205 & BIMA & $1-0$ & 28 & 3 & 1 \\
  SMC N83 & SEST & $2-1$,$1-0$ & 11,17 & 0.25 & 9 \\
  SMC LIRS36 & SEST & $2-1$ & 7 & 0.05 & 10 \\
  SMC LIRS49 & SEST & $2-1$ & 7 & 0.05 & 10 \\
  LMC N159 & SEST & $2-1$ & 6 & 0.25 & 11 \\
  NGC~1569 & PdBI & $2-1$,$1-0$ & 22,45 & 1.6 & 8 \\
  NGC~2976 & BIMA & $1-0$ & 94 & 3 & 2 \\
  NGC~3077 & BIMA & $1-0$ & 41 & 3 & 3 \\
  NGC~3077 & OVRO & $1-0$ & 41 & 1.3 & 6 \\
  NGC~4214 & BIMA & $1-0$ & 80 & 3 & 3 \\
  NGC~4214 & OVRO & $1-0$ & 64 & 1.3 & 7 \\
  NGC~4449 & BIMA & $1-0$ & 117 & 3 & 3 \\
  NGC~4605 & BIMA & $1-0$ & 109 & 3 & 4 \\
\cutinhead{Disk Galaxies}
Milky~Way & FCRAO & $1\to 0$ & $3-11$ & 0.65 & 12 \\
M~31 & BIMA & $1 \to 0$ & 27 & 2 & 14 \\
M~33 & BIMA & $1 \to 0$ & 30 & 2 & 13  \\
\enddata

\tablerefs{(1) \citet{YOUNG01}; (2) \citet{SIMON03}; (3) This Work;
  (4) \citet{BOLATTO02}; (5) \citet{WALTER03},\citet{LEROY06}; (6)
  \citet{WALTER02}; (7) \citet{WALTER01}; (8) \citet{TAYLOR99}; (9)
  \citet{BOLATTO03}; (10) \citet{RUBIO93a}; (11) \citet{BOLATTO00};
  (12) \citet{SANDERS86}; (13) \citet{ROSOLOWSKY03}; (14)
  \citet{ROSOLOWSKY07}}

\end{deluxetable}
\end{center}

\begin{center}
\begin{deluxetable}{lccccc}
\tabletypesize{\scriptsize}
\tablecolumns{6} 
\tablewidth{0pt} 
\tablecaption{\label{GALAXYTABLE}
    Galaxy Properties} \tablehead{ \colhead{Galaxy} &
    \colhead{Morph.} & \colhead{Dist.} & \colhead{$M_{B}$} &
    \colhead{Met.} & \colhead{Ref.}} 
\startdata
  IC~10 & Irr/BCD & 0.95 & $-16.7$ & 8.2 & 6,19\\
  NGC~185 & dSph/dE3 & 0.63 & $-14.7$ & 8.2 & 1,17 \\
  NGC~205 & E5 & 0.85 & $-15.9$ & 8.6 & 1,18 \\
  SMC & Sm & 0.061 & $-16.7$ & 8.02 & 7,20\\
  LMC & Sm & 0.052 & $-18.0$ & 8.43 & 7,20\\
  NGC~1569 & Irr & 2.2 & $-17.3$ & 8.19 & 2,16\\
  NGC~2976 & Sc & 3.45 & $-17.4$ & 8.7 & 3,13 \\
  NGC~3077 & Irr & 3.9 & $-17.5$ & 8.85 & 3,11 \\
  NGC~4214 & Irr & 2.94 & $-17.2$ & 8.23 & 4,12\\
  NGC~4449 & Irr & 3.9 & $-18.0$ & 8.32 & 5,15\\
  NGC~4605 & Sc & 4.26 & $-17.9$ & 8.69 & 3,14\\
\cutinhead{Large Galaxies}
Milky Way & SB & 0.008 & $-21.4$ & 8.7 & 10\\
M31 & Sb & 0.79 & $-21.1$ & 8.7 & 8,21 \\
M33 & Scd & 0.84 & $-18.9$ & 8.4 & 9,21\\
\enddata

\tablerefs{(1) \citet{RICHER95}; (2) \citet{KOBULNICKY97}; (3) This
work. See \S\ref{dwarfgal}; (4) \citet{KOBULNICKY96}; (5)
\citet{SKILLMAN89}; (6) \citet{LEQUEUX79}; (7) \citet{DUFOUR84}; (8)
\citet{PILYUGIN04}; (9) \citet{ROSOLOWSKYSIMON07}; (10)
\citet{BAUMGARTNER06}; (11) \citet{SAKAI01}; (12) \citet*{APELLANIZ02};
(13) \citet{SIMON03}; (14) \citet{BOLATTO02}; (15) \citet{HUNTER05};
(16) \citet{ISRAEL88}; (17) \citet{MARTINEZ98}; (18)
\citet{SALARIS98}; (19) \citet{HUNTER01}; (20) \citet{KELLER06}; (21)
\citet{KENNICUTT98}}

\tablecomments{Distances are in Mpc. Luminosities are corrected for
Galactic extinction. Metallicities are expressed as $12+\log{O/H}$.}
\end{deluxetable}
\end{center}

\subsection{Dwarf Galaxies}
\label{dwarfgal}

In this section we briefly summarize the sources and datasets
used. For more in--depth descriptions we point the reader to the
original references.

\citet{YOUNG01} used the BIMA interferometer to map CO $J=1\rightarrow0$
emission in the Local Group dwarf elliptical galaxies NGC~185 and NGC~205.
These galaxies, which are satellites of M~31, are unusual early-type dwarfs
with CO emission, while the rest of our sample consists of late-type dwarf
spiral and irregular galaxies.

\citet{SIMON03} presented a kinematic study based on BIMA observations
of CO $J=1\rightarrow0$ emission from NGC~2976, a dwarf spiral galaxy
in the M~81 group. These observations are considerably deeper than
those from the BIMA Survey of Nearby Galaxies
\citep[SONG;][]{HELFER03}.  The metallicity for this galaxy is
uncertain, but the polycyclic aromatic hydrocarbon abundance and
dust-to-gas ratios derived by \citet{DRAINE07} suggests it is similar
to Galactic.

\citet{WALTER02} mapped CO $J=1\rightarrow0$ emission from NGC~3077, a
member of the M~81 group currently undergoing a dramatic interaction
with M~81 and M~82, using the OVRO millimeter interferometer. This
galaxy was also observed by BIMA during the same period, and those
observations appear here for the first time. We present GMC property
measurements from both data sets. We derive the metallicity for this
galaxy using the integrated nebular fluxes of the bright oxygen lines
by \citet{MOUSTAKAS06}, employing the calibration of \citet{MCGAUGH91}
as parametrized by \citet*{KOBULNICKY99}.  The metallicity is
consistent with that derived using the nitrogen calibration by
\citet{KEWLEY02}.

\citet{WALTER01} mapped CO $J=1\rightarrow0$ emission from NGC~4214, a
nearby dwarf irregular galaxy, using the OVRO millimeter
interferometer. This galaxy was also observed by BIMA, and here we use
those data for the first time.

BIMA mapped CO $J=1\rightarrow0$ emission from the nearby dwarf
irregular NGC~4449. The data appear here for the first time.

\citet{BOLATTO02} presented CO $J=1\rightarrow0$ observations of
NGC~4605, a nearby isolated dwarf galaxy, and used them to study the
kinematics of this galaxy. The metallicity for this galaxy was derived
using the same procedure as for NGC~3077.

\citet{WALTER03} mapped CO $J=1\rightarrow0$ emission from most of the
molecular clouds in the Local Group dwarf irregular IC~10 using the
OVRO millimeter interferometer. These data were analyzed in detail by
\citet{LEROY06}, where maps of individual GMCs are presented. IC~10 is
a Local Group irregular comparable in mass to the Small Magellanic
Cloud but presently undergoing a burst of vigorous star formation.

\citet{TAYLOR99} used the PdBI to map both CO $J=2\rightarrow1$ and
$J=1\rightarrow0$ emission from the nearby dwarf irregular
NGC~1569. This galaxy, which is about the mass of the Small Magellanic
Cloud, experienced a starburst period that ended $\sim 5$~Myr ago. 

\citet{BOLATTO03} presented high signal-to-noise mapping
of the N83 region in the Small Magellanic Cloud (SMC) using the SEST
telescope. They observed both the CO $J=2\rightarrow1$ and
$J=1\rightarrow0$ transitions. N83/N84 is a bright star forming region
in the eastern wing of the Small Magellanic Cloud.

\citet{BOLATTO00} used SEST to map CO $J=2\rightarrow1$ emission in
the N159 region of the Large Magellanic Cloud (LMC). This region lies just
south of the bright star forming region 30 Doradus, and comprises
clouds in different stages of evolution.

\subsection{Large Galaxies}
\label{disks}

In this study we use GMCs properties measured in the spiral disk
galaxies of the Local Group as comparison and control datasets.  As
discussed in the introduction, several studies have established the
scalings for Galactic molecular clouds
\citep{LARSON81,SOLOMON87,ELMEGREEN96,HEYER01,HEYER04,ROSOLOWSKY08}. Recent
interferometric work has shown that the GMCs in the other disk
galaxies in the Local Group (M~31 and M~33) follow the same scalings
\citep{WILSON90,ROSOLOWSKY03,ROSOLOWSKY07,SHETH08}.

The observations of GMCs in M~33 employed here were made using the
BIMA millimeter interferometer and are presented by
\citet{ROSOLOWSKY03}.  That study analyzed 20 fields in that galaxy
with a resolution $\theta\sim 6''\sim 20\mbox{ pc}$, targeting sources
identified in the all-disk survey of \citet{ENGARGIOLA03}.  The M~31
GMCs belong to a region along the northwest spiral arm of this galaxy,
also observed by BIMA \citep{ROSOLOWSKY07}. This region was chosen for
its favorable observing geometry, strong dust lanes and active star
formation.  Both studies concluded that the GMCs in these disks were
indistinguishable in their aggregate properties from the clouds in the
inner Milky Way.  These observations constitute the most complete data
sets for these galaxies, have sensitivity and resolution comparable to
the dwarf galaxy data, and have been analyzed in an identical fashion.

For the Milky Way we employ the catalog of GMC properties by
\citet{SOLOMON87}. This catalog consists of size, line width, and
luminosity measurements for 273 Galactic GMCs and constitutes a
representative data set of Milky Way GMC properties. Like us (see
\S\ref{method}), \citeauthor{SOLOMON87} use moments to measure cloud
properties and isosurface-based methods to identify individual GMCs.
Their study, however, does not make an explicit correction for the
effects of sensitivity (they use boundary isosurfaces of brightness
temperature T$_B=4$ to 7~K). It also uses an outdated Galactocentric
distance for the Sun (10 kpc). It is unclear how to best correct
correct their catalogued cloud properties to bring them into closer
agreement with our methodology. We use here the original numbers, and
rely on the M~31 and M~33 data as a consistency check. As is apparent
in Figures \ref{size_linewidth}, \ref{lumlw}, and \ref{lumsize}, these
data are in good agreement with the standard Larson relations derived
for the Milky Way sample.

\section{Methodology}
\label{method}

The challenge of any study that uses different instruments to study
galaxies with a range of intrinsic properties and distances is to
control the systematic biases introduced by the differences in the
datasets.  We measure cloud properties using the algorithm described
in detail by \citet{ROSOLOWSKY06} and summarized here. This method
measures luminosities, line widths, and sizes of GMCs in a manner that
minimizes the biases introduced by signal-to-noise and
resolution. These biases are the Achilles heel of comparisons of GMC
measurements across different galaxies obtained by different
telescopes.  \citeauthor{ROSOLOWSKY06} proposed that moment
measurements combined with beam deconvolution and extrapolation
represent a robust way to compare heterogeneous observations of
molecular clouds. The algorithm is part of a package called {\tt
CPROPS} implemented in IDL and is available upon request. The
remainder of this section summarizes the approach step-by-step.

The robustness, reliability, and consistency of the {\tt CPROPS}
methodology was tested by \citet{ROSOLOWSKY06} for a globally
applicable set of algorithm parameters.  They found the algorithm
performed very well for data with signal-to-noise ratios (S/N)
$\gtrsim 10$, but cautioned that measurements of the properties of
GMCs observed with low S/N remain uncertain, particularly when using
the default set of algorithm parameters.  Unfortunately, CO is faint
in dwarf galaxies, which makes it difficult to attain the optimal
combination of S/N and spatial resolution.  Therefore, only a few of
the datasets discussed here meet the S/N$\geq10$
criterion. Nevertheless, the data assembled here represent the best
observations of GMCs in dwarf galaxies to date.  As a consequence of
the low S/N data, we made minor adjustments described below to the
standard {\tt CPROPS} parameters to improve the reliability of our results.

\subsection{Identify Signal In Each Data Set}

First, we identify regions of significant emission in each data set.
For most of our data these are defined as regions that contain pixels
which satisfy the condition of having 3 consecutive velocity channels
above $4\sigma$ significance (i.e. a high significance ``core''). We
extend these regions to include all adjacent data which has at least 2
consecutive channels above $1.5\sigma$ significance. For a few data
sets this did not yield the best results and we adjusted either the
core or the edge conditions: a higher edge threshold was used in the
high S/N N83 and N159 data sets; lower core thresholds were needed for
the low S/N NGC~185 and NGC~205 data sets; a few other minor
adjustments were also made. These changes yielded stable properties for
clouds of low S/N, and suppressed diffuse emission blending clouds
together in the extremely high S/N case.  The signal identification
step, often referred to as ``masking'', can have a large impact on
derived cloud properties and is difficult to motivate physically,
especially in the low S/N regime common to much of our data. Thus, by
necessity this step introduces a certain amount of subjectivity in the
analysis.

\subsection{Apportion Emission Into Clouds}

Once signal is identified, data is apportioned into individual clouds
following \citet{ROSOLOWSKY06}. In all cases, we use the ``physical
priors'' described in their appendix, which set the parameters of the
decomposition to common values motivated by the properties of Galactic
GMCs. This entails decomposing emission into clouds using effective
spatial and velocity resolutions of 15~pc and 2~km~s$^{-1}$ and
ignoring even high-contrast substructure below these scales. When it
is not possible to achieve these resolutions because of limitations in
the data, we use the actual spatial or velocity resolution of the
observations. When two potential clouds share an isosurface, the
algorithm adopts a conservative approach to separating them.  The
clouds are considered separate only if all of the following conditions
are met: 1) each is large enough to measure meaningful properties, 2)
each shows a minimum contrast between peak and edge, and 3) the
decision to break the isosurface into multiple clouds has a
significant effect on the derived cloud
properties. \citet{ROSOLOWSKY06} showed this approach to be
conservative and robust in the presence of noise, minimizing the
identification of low-signal, marginally resolved (and thus often
spurious) clouds.

\subsection{Use Moments to Measure Cloud Properties}

After identifying the individual clouds, we calculate their size,
velocity dispersion, and CO luminosity. We determine the major and
minor axis of the cloud using principal component analysis and measure
the second moments of emission distribution along these axes. The RMS
size of the cloud ($\sigma_r$) is the geometric mean of these two
moments.  The velocity dispersion ($\sigma_v$) is measured from the
second moment along the velocity axis, and the luminosity via the
zeroth moment (sum) over the cloud.

At finite signal-to-noise the second moment will underestimate the
true cloud size, introducing a sensitivity bias. Although for Galactic
data it is possible to measure reliable properties by adopting a fixed
brightness boundary for clouds \citep[e.g.,][]{SOLOMON87}, it would be
impossible to compare measurements of different galaxies without
removing this bias. We avoid the sensitivity bias by measuring the
size, velocity, and luminosity as a function of intensity isosurface
and extrapolating this to the case of infinite signal-to-noise.  The
size and line width of the cloud are extrapolated linearly, the
luminosity is extrapolated quadratically \citep[see][for a
justification of the extrapolation orders]{ROSOLOWSKY06}. We discussed
in \S\ref{disks} the corrections that we apply to the Galactic
sample to bring it into agreement with our methodology.
  
It is often the case for extragalactic observations that the size and
line width of a GMC are comparable to the angular and velocity
resolution of the telescope. We correct for this by ``deconvolving''
both the telescope beam and the velocity channel profile. We do so by
subtracting their values from the extrapolated moment measurements in
quadrature. Clearly this step does not come for free in
signal-to-noise. Size measurements of clouds that are only marginally
resolved suffer from substantial uncertainty, and this is accounted
for in the error estimates.

\subsection{Derive Physical Quantities}

From the moment measurements we convert to physical units --- radius,
luminous mass, virial mass.  The velocity dispersion is given
directly by the moment.  

We convert the RMS size to the spherical radius of the cloud using the
factor ${3.4}/{\sqrt{\pi}}$, so that radii in this paper follow the
\citet{SOLOMON87} definition: $R=1.91 \sigma_r$. This factor can be
motivated by considering a constant density spherical cloud and
comparing its RMS size to its radius.

We compute luminous masses, $M_{lum}$, using the formula

\begin{equation}
M_{lum}=1.1\times10^4\, S_{\rm CO} \,D^2\ {\rm M}_\odot,
\label{mlum}
\end{equation}

\noindent where $S_{\rm CO}\equiv\int{\rm I_{\rm CO}}\,d\Omega~dv$ is the
integrated $^{12}$CO flux of the molecular cloud measured in
Jy~\kmpers, $D$ is the distance to the source in Mpc, and the
coefficient corresponds to our adopted conversion factor $\xco = 2
\times 10^{20}$ \xcounits, and includes the contribution of the
cosmological He abundance to the total mass.  Equivalently, for a CO
luminosity $L_{\rm CO}$ expressed in \Kkmpers~pc$^{2}$,

\begin{equation}
M_{lum}=4.5\, L_{\rm CO}\ {\rm M}_\odot. 
\label{mlum2}
\end{equation}

We compute the virial masses under the assumption that each cloud is
spherical and virialized with a density profile of the form $\rho
\propto r^{-1}$. This assumption is made for consistency with previous
work, and we cannot assess its validity with the data in hand. Thus
the virial mass is given by the formula \citep{SOLOMON87}

\begin{equation}
M_{vir} = 1040\, \sigma_v^2 \, R\ {\rm M}_{\odot},
\label{virialmass}
\end{equation}

\noindent where $\sigma_v$ is the cloud velocity dispersion in
\kmpers, $R$ is the spherical cloud radius in pc, and $M_{vir}$ is the
cloud virial mass in M$_\odot$. Implicit in this equation is the
assumption that the virial parameter equals one.

Other works, in particular those focused on Milky Way clouds, have
treated clouds as ellipsoids and used the corresponding form of the
virial theorem \citep[e.g.][]{BERTOLDI92}, instead of the simplified
Equation \ref{virialmass}.  Because of the low resolution and S/N of
extragalactic observations, this approach is untenable. Under these
conditions, deconvolution becomes very difficult and the precise shape
of the source is hard, if not impossible, to recover.  We note that
the more complex measurements of cloud dimensions reduce to the
\citet{ROSOLOWSKY06} RMS size parameter for use in the
\citet{BERTOLDI92} formulation of the the virial theorem.

To relate the mass surface density, $\Sigma_{\rm GMC}$, to the coefficient
of the {\em size-line width} relation, $\sigma_v=C\,R^{0.5}$, we
simply divide Equation \ref{virialmass} by the area of the cloud,
$\pi\,R^2$, to obtain

\begin{equation}
\Sigma_{\rm GMC} = 331\, C^2,
\label{surfacedensity}
\end{equation}

\noindent where the coefficient $C$ is in units of \kmpers~pc$^{-0.5}$ and
the surface density is expressed in M$_\odot$~pc$^{-2}$.

\subsection{Estimate Uncertainties}

The errors quoted in this paper are derived from bootstrapping the
data in each cloud and re-deriving its properties. This yields a
realistic assessment of the uncertainty {\em once signal has been
apportioned into clouds}. We also include a 25\% gain uncertainty in
the luminosity, typical of the uncertainty in the flux calibration at
millimeter wavelengths. In high S/N cases, e.g. our N~83 and N~159
data sets, bootstrapping yields very low uncertainties. In these
cases, virtually all of the uncertainty in the derived cloud
properties rests with the apportionment of emission into clouds. This 
uncertainty is not included in our results.

\section{Results}
\label{results}

\subsection{Derived Cloud Properties}
\label{dercloudprop}

Table \ref{BIMATAB} presents our measurements of GMC properties in
dwarf galaxies (available in full in the electronic edition). In order
the columns list: (1) the galaxy that contains the GMC; (2) a number
identifying the cloud; (3) the transition, CO $J=2\rightarrow1$ or
$1\rightarrow0$, being observed; (4) the right ascension and (5)
declination of the intensity-weighted cloud center; (6) the mean LSR
velocity of the cloud in km s$^{-1}$; (7) the log of the cloud
radius, in parsecs; (8) the log of the line of sight velocity
dispersion of the cloud, in km s$^{-1}$; (9) the log of the
luminous mass from Equation \ref{mlum}; (10) the log of the
virial mass from Equation \ref{virialmass}; and (11) the peak
brightness temperature in the cloud, in Kelvin.  Uncertainties from
bootstrapping are listed with each measurement.

Table \ref{restab} summarizes some of our results on a per galaxy
basis. In order the columns list: (1) galaxy name; (2) number of
measured GMC sizes in either CO transition; ($3-4$) average and range
of GMC masses, in units of 10$^5$ M$_\odot$; ($5-6$) error-weighted
average and range of the logarithm of \xco, in units of our adopted
Galactic value; ($7-8$) error-weighted average and range of the
coefficient of the {\em size-line width} relation, $\sigma_v
R^{-0.5}$; and (9) corresponding average GMC surface density from
Equation \ref{surfacedensity}.

We checked for correlations between parameters in Table \ref{restab}
and distance or resolution, and found none obvious. The major bias
present, which is unavoidable, is that there is a correlation between
our minimum virial mass and distance that is simply due to
sensitivity. In that sense we sample different ranges of GMC mass in
different galaxies. Considerations such as the nonlinearity of the
{\em mass-luminosity} relation, for example, become important in the
Magellanic Clouds where the cloud masses are small.

\subsection{The Extragalactic Size-Line Width Relation}
\label{sizelw}

Of the three Larson relations perhaps the most fundamental one is that
between cloud size and line width (Equation \ref{size_linewidth_eq}),
as it is mostly (but not entirely) independent of the details of the
excitation of the tracer molecule.  The initial studies already
pointed out the connection between the {\em size-line width} relation
and expectations from the Kolmogorov theory for subsonic turbulence in
incompressible fluids \citep{LARSON79,LARSON81}. Later measurements in
the inner Galaxy allowed a better determination of the relationship,
showing that the exponent is not 1/3 as it would be expected for a
pure Kolmogorov cascade, but rather the steeper $0.5\pm0.05$ used in
Equation \ref{size_linewidth_eq} \citep{SOLOMON87}. An index of $0.5$
is consistent with expectations for compressible supersonic
magnetohydrodynamic turbulence \citep*[also known as Burger's
turbulence,][]{VESTUTO03,BRUNT04}. \citet{HEYER04} show that this
relationship extends also within GMCs, which is consistent with the
multi-scale turbulence interpretation of the observed line widths.
The authors argue that because the {\em size-line width}
relation is well-defined it implies the invariance and universality of
the turbulent scaling relations.  For clouds in virial equilibrium
$M\sim R \sigma_v^2$, thus the observed {\em size-line width} relation
implies $M\sim R^2$, or equivalently that all clouds have similar
surface densities.

Recent studies of cloud properties in the outer Galaxy have shown that
the {\em size-line width} relation breaks down for very small clouds.
In these objects the velocity dispersion $\sigma_v$ becomes a constant
independent of their size \citep{HEYER01}. These authors find that
outer disk clouds smaller than $\sim7$ pc in radius have an almost
constant velocity dispersion $\sigma_v\sim0.8$ \kmpers, while clouds
larger than $\sim7$ pc follow the usual scaling. \citeauthor{HEYER01}
interpret this fact in terms of the dynamical equilibrium state of the
small clouds. These clouds have a virial parameter
$\alpha=M_{vir}/M_{lum}$, representing their ratio of kinetic to
gravitational energy, that is $\alpha\geq1$ for masses below
$M\sim1.3\times10^4$ M$_\odot$. Thus these small clouds have an excess
of kinetic energy over their self-gravity, are not in virial
equilibrium, and in order to be bound they have to be confined by an
external pressure of order $P/k\sim10^4$ \percmcu~K. The precise
location of the break in the {\em size-line width} relation is then
likely a function of the local pressure of the cloud
environment. Similarly, analysis of the $^{13}$CO emission from GMCs
in the inner 3 kpc of the Milky Way reveals a less steep {\em
size-line width} relation than that obtained from $^{12}$CO studies,
$\sigma_v\sim R^{0.15-0.30}$ \citep{SIMON01}. These authors conclude
that although cloud complexes are bound, many of the clumps identified
in their study are not gravitationally bound, probably due in part to
the methodology of their clump identification algorithm. Finally,
\citet{OKA01} measured the GMC properties near the Galactic Center,
finding that those clouds have line widths a factor of 3 to 5 times
larger than similar clouds in the disk and suggesting that this is due
to the large pressure in this region. These studies show that there
are observed departures from the traditional {\em size-line width}
relation in the Milky Way, in the direction of larger line widths for
a given size.

\begin{figure*}
\plotone{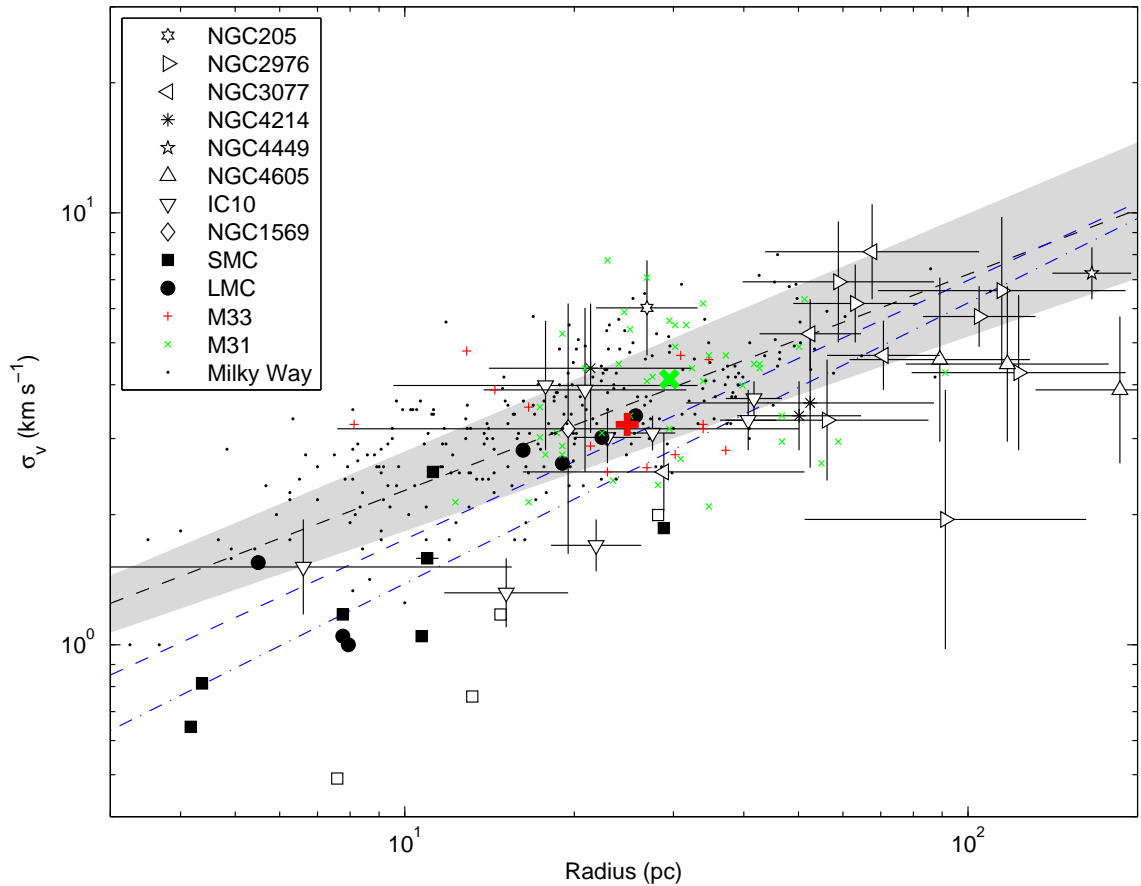}
\caption{Extragalactic {\em size-line width} relation. The different
symbols and the associated error bars indicate the measurements of
molecular clouds in different galaxies. The black dots correspond to
the sample of Milky Way clouds by \citet{SOLOMON87}. The open and
filled symbols indicate measurements based on CO $(1-0)$ and $(2-1)$
observations respectively. The dashed black line and the gray area
indicate the standard relationship for the Milky Way and its
uncertainty, $\sigma_v=(0.72\pm0.07) R^{0.5\pm0.05}$.  The dashed blue
line illustrates the fit to the extragalactic GMC sample represented
by Equation \protect\ref{exgal_linewidth_size}. The dash-dotted blue line
indicates the fit to the dwarf galaxies only. The large color symbols
indicate the median centroids for the GMCs in M~31 and M~33, our
control sample. Dwarf galaxies are mostly compatible with the Galactic
{\em size-line width} relation, with a tendency to deviate in the
direction of smaller velocity dispersions for a given cloud size. In
this regard the most discrepant points belong to the Small Magellanic
Cloud and IC~10, some of the lowest metallicity galaxies in the
sample. This is not, for the most part, a bias introduced by our
methodology as the M~31 and M~33 points appear to follow the Galactic
relationship.\label{size_linewidth}}
\end{figure*}

Figure \ref{size_linewidth} shows our results for the ensemble of
extragalactic clouds. Remarkably, dwarf galaxy cloud measurements are
broadly compatible with the Milky Way {\em size-line width} relation,
represented by the gray area and the small dots from the sample of
\citet{SOLOMON87}. Detailed inspection of the data reveals, however,
that there is a tendency for GMCs in dwarf galaxies to fall
preferentially under the Galactic relationship. This is most apparent
in a few clouds in the LMC, IC~10, and most clouds in the SMC, which
are significantly inconsistent with the Milky Way. Note that these
departures occur in the direction opposite from those described in the
previous paragraph. While we may suspect that cloud measurements
yielding sizes $R\gtrsim100$~pc are affected by blending (such as
those on the right side of the diagram), we have no reason to think
that there are problems with the LMC, SMC, or IC~10, where we have
some of the best spatial resolutions in the dataset.  Furthermore, the
control sample of M~31 and M~33 GMCs is entirely consistent with the
Galactic {\em size-line width} relation and does not suffer from the
same offsets. We discuss the particular situation of the SMC, which
exhibits the most convincing departures, in \S\ref{smc}.  Note that
\citet{RUBIO93b} arrived at a different conclusion in their study of
the SMC. Because the GMC samples do not substantially overlap it is
difficult to assess the source of the discrepancy. We attribute it
mostly to the differences in the analysis methodologies, albeit it is
possible that the samples are intrinsically different (perhaps due to
different radiation environments).

Although it is not entirely correct to derive a unique {\em size-line
width} relation for the ensemble of extragalactic data, we have
determined a best fit relation. In order to avoid driving the result
with the small formal error bars in the Magellanic Clouds, and to
better represent the scatter in the data, we increase the errors in
quadrature in both variables until we obtain a best fit $\chi^2$ value
of unity. The resulting relation, illustrated by the dashed blue line
in Figure \ref{size_linewidth}, is

\begin{equation}
\sigma_v\approx0.44^{+0.18}_{-0.13}\, R^{0.60\pm0.10}\ {\rm km~s^{-1}}~,
\label{exgal_linewidth_size}
\end{equation}

\noindent where the data are fit in the log--log plane with
simultaneous errors in both variables, and the error bars are derived
using a bootstrapping technique. If we only use the dwarf galaxy data,
the coefficient and the exponent are $0.31^{+0.09}_{-0.07}$ and
$0.65\pm0.07$ respectively. 

If the {\em size-line width} relation is interpreted in terms of
virialized self-gravitating clouds supported mainly by turbulence, the
compatibility between the Galactic relation and our measurements
implies that most GMCs in dwarf galaxies have surface densities not
very different from Galactic GMCs, for which $\Sigma_{\rm GMC}\approx170$
\msunperpcsq.  Extragalactic GMCs that fall under the Galactic {\em
size-line width} relation would have lower surface and volume
densities than corresponding clouds in the Milky Way.  The small-size,
low velocity dispersion end of the dash-dotted line in Figure
\ref{size_linewidth} corresponds to a drop of approximately a factor
of 2 in the velocity dispersion, which translates to a factor of 4
decrease in surface density with respect to the Galactic value. This
surface density may be appropriate for the most extreme galaxies in
the sample (c.f., Table \ref{restab}). If we fit a model that is
proportional to $R^{0.5}$, the sample of dwarf galaxies is consistent
with an average surface density $\Sigma_{\rm GMC}\sim85$ \msunperpcsq.

An alternative possibility is that the virial parameter, a measure of
the ratio of kinetic to gravitational energy of the cloud
\citep{BERTOLDI92}, is different for these objects.  The virial
parameter, $\alpha$, can be expressed in terms of the velocity
dispersion following Equation 27 in \citet{MCKEE07} to obtain
$\sigma_v=\sqrt{\alpha\Sigma_{\rm GMC}}\,R^{-0.5}$. Thus, a low
coefficient in the {\em size-line width} relation can be attributed to
a lower than Galactic virial parameter. Bound clouds in the Milky Way
have $\alpha\sim1$, while pressure-confined clouds exhibit
$\alpha\gg1$ --- a cloud in equilibrium with $\alpha<1$ requires some
additional type of support, presumably provided by magnetic fields. On
the spatial scales studied here, such support is usually unimportant
in the Milky Way.

\subsection{The Luminosity-Line Width and Luminosity-Size relations}
\label{luminosity_relations}

\begin{figure*}
\begin{center}
\plotone{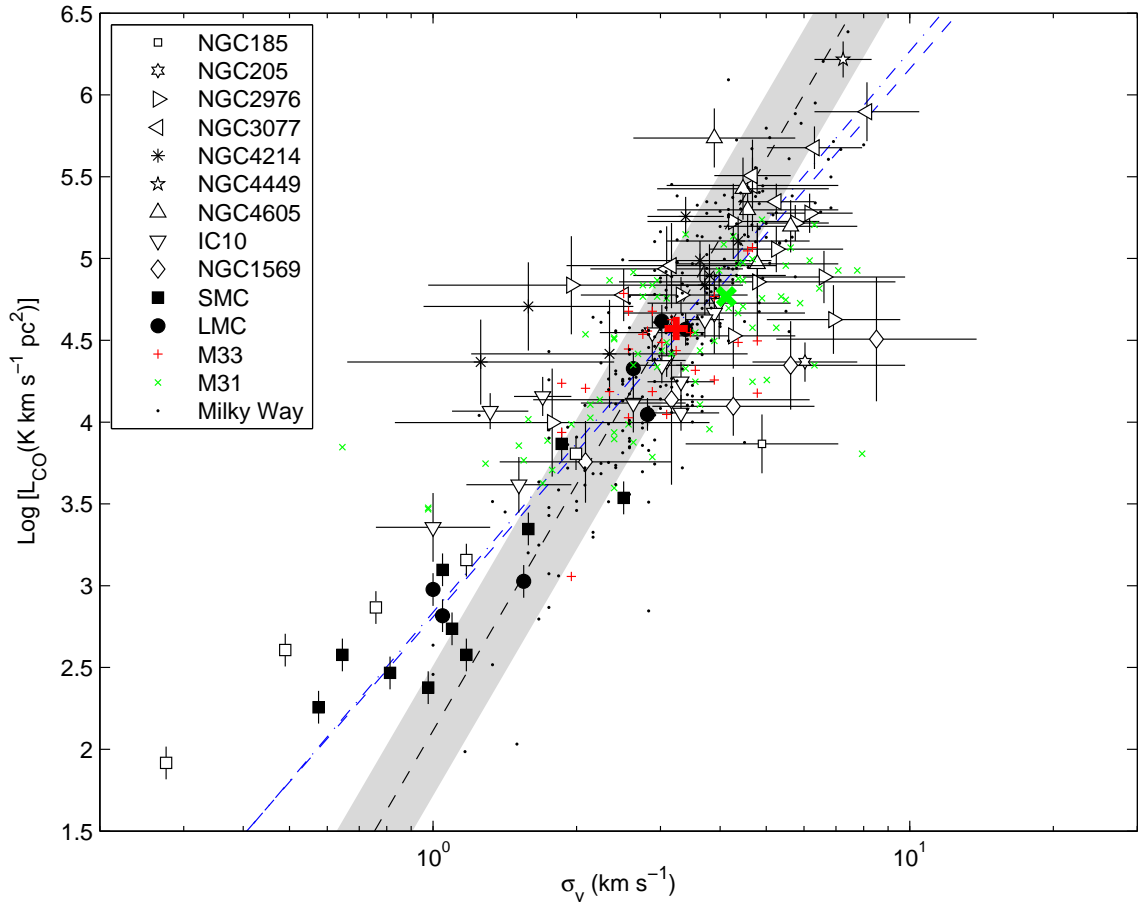}
\end{center}
\caption{Extragalactic {\em luminosity-line width} relation. As in
Fig. \protect\ref{size_linewidth}, the different symbols and the
associated error bars indicate the measurements of molecular clouds in
different galaxies, with open and filled symbols indicating
measurements based on CO $(1-0)$ and $(2-1)$ observations
respectively. The black dots correspond to the sample of Milky Way
GMCs by \citet{SOLOMON87}, and the gray regions to their $1\sigma$
dispersion around the dashed lines, which follow Equations
\protect\ref{lum_lw_eq} and \protect\ref{lum_size_eq}. The dashed blue
line illustrate the fits to the extragalactic GMCs.  The dash-dotted
blue line corresponds to the results of the fits for the dwarf galaxy
data only.  The large color symbols indicate the median centroids of
the clouds that have $R$ measurements for our control sample of GMCs
in M~31 and M~33. \label{lumlw}}
\end{figure*}

\begin{figure*}
\begin{center}
\plotone{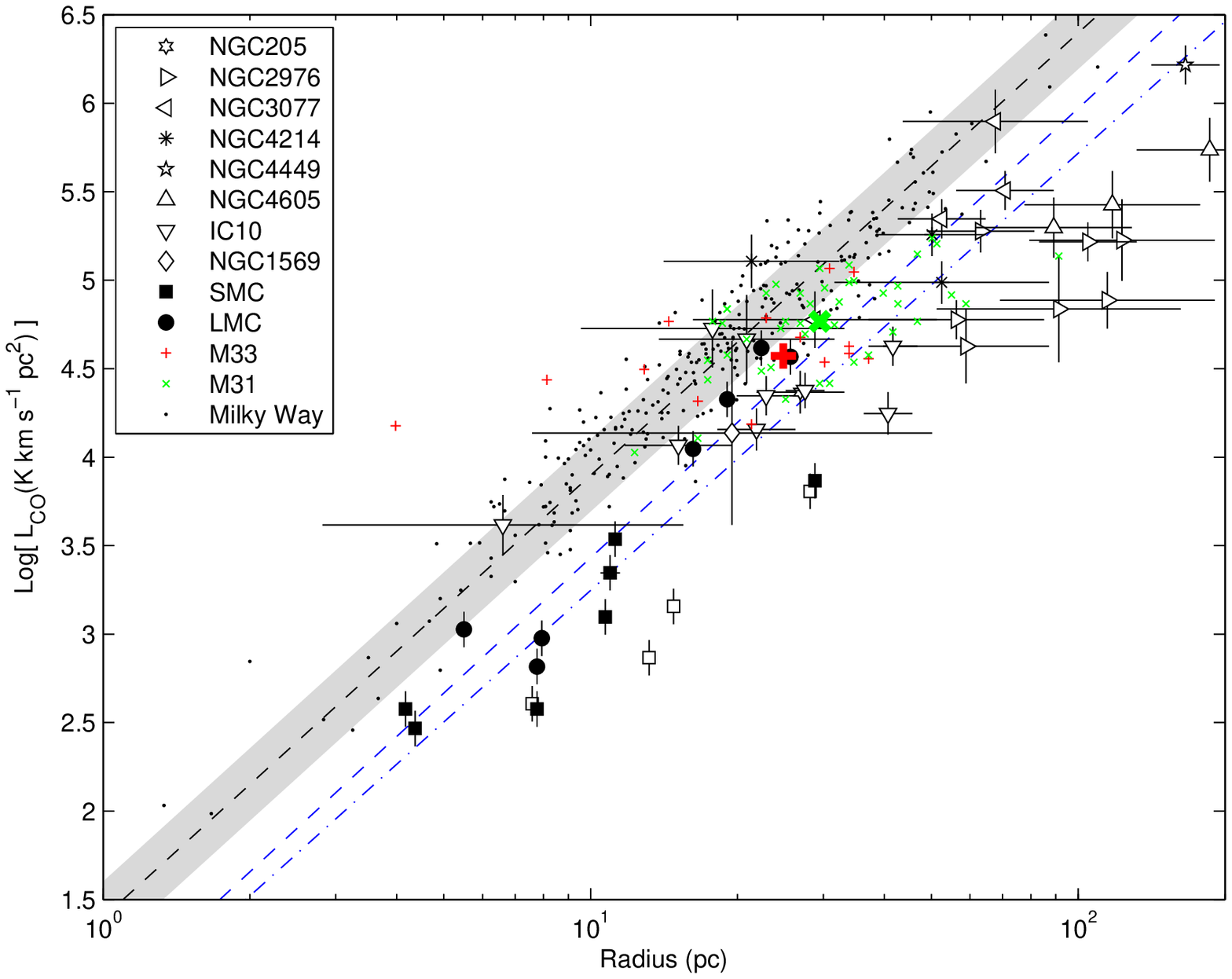}
\end{center}
\figcaption{Extragalactic {\em luminosity-size} relation. 
Symbols and labels as in Fig. \ref{lumlw}. \label{lumsize}}
\end{figure*}

Figures \ref{lumlw} and \ref{lumsize} show the relations between $^{12}$CO
luminosities and either velocity dispersion or cloud radius. These
luminosities can be related to the corresponding luminous mass
$M_{lum}$ for a Galactic \xco\ using Eqs. \ref{mlum} and
\ref{mlum2}. The steep dependency on $\sigma_v$ in Equation
\ref{lum_lw_eq} is a direct consequence of the observed {\em size-line
width} relation in combination with the assumption of virial
equilibrium. The coefficient in front of this equation reflects a
combination of the area filling fraction by CO clumps within the
resolved cloud, and the kinetic temperature of the gas itself.

\begin{figure*}
\plotone{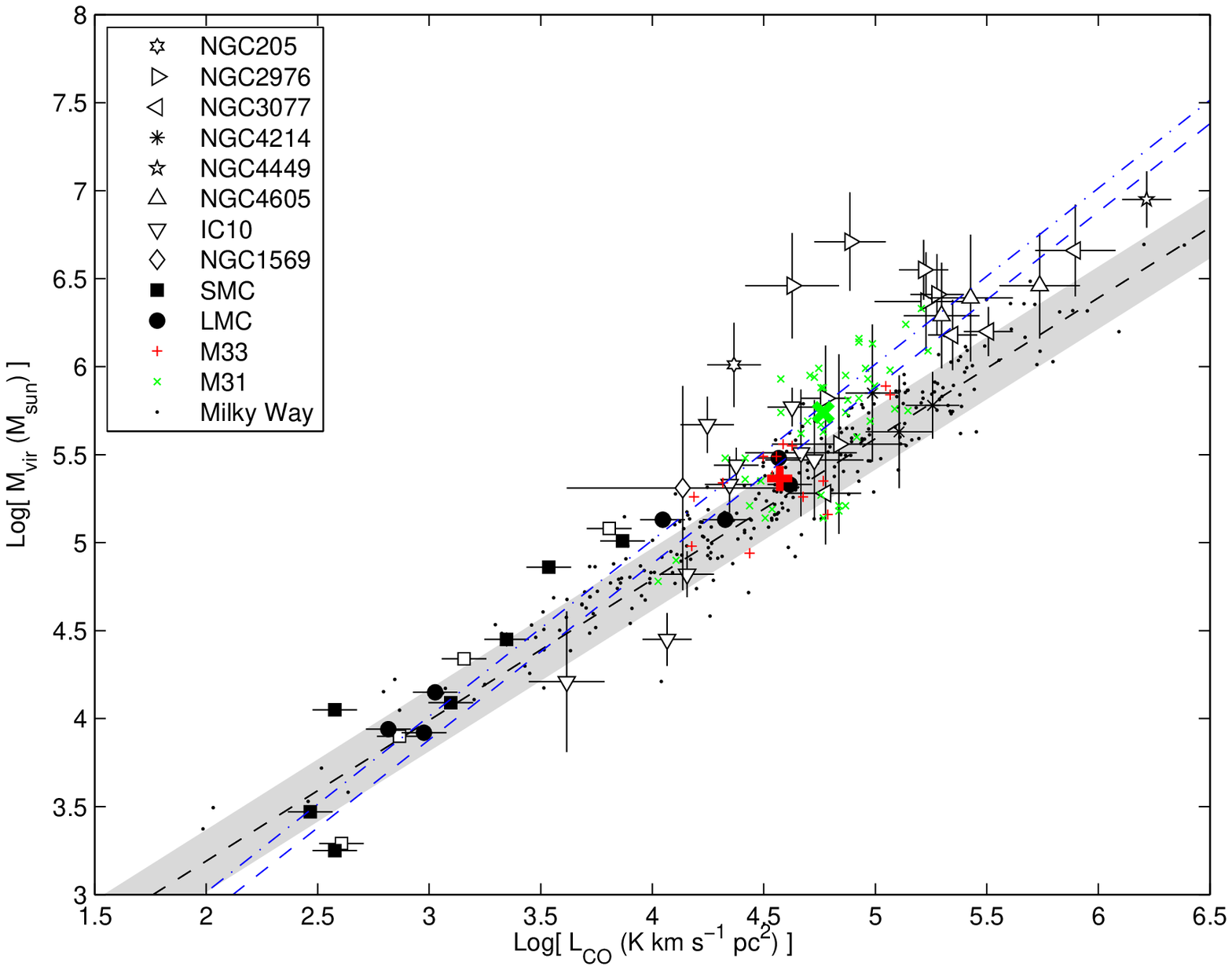}
\caption{The relationship between CO luminosity, $L_{\rm CO}$, and
virial mass, $M_{vir}$, for Galactic and extragalactic GMCs. As in
previous figures white and black symbols indicate measurements based
on CO $1-0$ and $2-1$ respectively. The dots correspond to the sample
of Milky Way GMCs by \citet{SOLOMON87}. The dashed black line
illustrates equation \protect\ref{lco_mvir_eq}, with the gray region
corresponding to the Galactic $1\sigma$ dispersion around the line
($\approx50\%$). The dashed and dash-dotted blue lines illustrate the
fits for all extragalactic GMCs and for the dwarf galaxy sample alone,
respectively.
\label{lco_mvir}}
\end{figure*}

Inspection of Figures \ref{lumlw} and \ref{lumsize} reveals that,
despite the variety of environments sampled and the consistently
lower-than-Galactic metallicity of our sources, there is good
agreement between the properties of extragalactic and Galactic
GMCs. Nonetheless, it is clear that there is an overall tendency for
clouds in dwarf galaxies to be underluminous for their size, and
clouds in the SMC tend to be also overluminous for their velocity
dispersion.  In other words, they tend to be larger and (in the case
of the SMC) more quiescent (as traced by their velocity dispersion)
than Milky Way GMCs of similar luminosity. The fit to all the
extragalactic data, conducted as in \S\ref{sizelw}, shows that the
{\em luminosity-line width} relation is

\begin {equation}
L_{\rm CO}\approx645^{+165}_{-132}\, \sigma_v^{3.35\pm0.19}\
\Kkmperspcsq,
\label{exgal_lum_lw_eq}
\end{equation}

\noindent where the shallower--than--Galactic slope is driven mostly by
the small SMC clouds. A similar analysis but including only the dwarf
galaxy data yields a coefficient $688^{+220}_{167}$ and an exponent
$3.43\pm0.22$, thus very consistent with the results when M~31 and
M~33 are included. The corresponding extragalactic {\em
luminosity-size} relation is

\begin{equation}
L_{\rm CO}\approx7.8^{+6.9}_{-3.7}\, R^{2.54\pm0.20}\ \Kkmperspcsq,
\label{exgal_lum_size_eq}
\end{equation}

\noindent where the corresponding coefficient and exponent when
fitting only the dwarf galaxy data are $6.0^{+3.8}_{-2.3}$ and
$2.47\pm0.17$. This shows that the data are completely consistent with
the Galactic slope, but the relation is noticeably displaced toward
lower luminosities for a given size.

Figure \ref{lumsize} suggests that, despite our efforts to correct for
them, there may be a remaining methodological offset in the cloud size
determination between the sample of \citet{SOLOMON87} and this study,
of order 30\% toward larger $R$ (equivalent to a factor of $0.5$
toward lower luminosities).  Unfortunately, establishing the origin of
these offset would require a full reanalysis of the original Galactic
survey.  With the quality of the data at hand it is difficult to
assert whether methodological or real offsets are driving the observed
shift toward larger cloud sizes, although it seems that the
likely methodological offsets are too small to explain the level of
discrepancy in many sources --- most notably the SMC.

Figure \ref{lco_mvir} illustrates the behavior of luminosity and
virial mass described by equation \ref{lco_mvir_eq} for our sample of
extragalactic GMCs.  \citeauthor{SOLOMON87} discuss this empirical
relation in some detail, showing that it is a natural outcome of the
combination of the {\em size-line width} relation with the assumption
of virial equilibrium when all clouds share similar kinetic
temperatures and area filling fractions in their resolved CO
emission. Despite the aforementioned tendency for clouds in dwarf
galaxies to be underluminous for their sizes, their virial masses are,
on average, only slightly underpredicted by their CO luminosity.  In
the particular case of GMCs in the SMC, because the observed
departures from the standard {\em luminosity-line width} and {\em
luminosity-size} relations occur in opposite directions, their larger
sizes and smaller velocity dispersions for a given luminosity
approximately cancel out in the virial mass calculation.  Thus most
reliable extragalactic clouds in our sample are remarkably compatible
with equation \ref{lco_mvir_eq}. The fit to the entire extragalactic 
sample yields

\begin{equation}
M_{vir}\approx7.6^{+3.9}_{-2.6}\, L_{\rm CO}^{1.00\pm0.04}\ {\rm M}_\odot, 
\label{exgal_lco_mvir_eq}
\end{equation}

\noindent while the fit to the dwarf galaxies only has a coefficient
of $10.3^{+5.8}_{-3.7}$, and an exponent $1.00\pm0.05$. Thus the
extragalactic and Galactic {\em mass-luminosity} relations are in
excellent agreement for GMCs in the mass range $10^3-10^5$ M$_\odot$.
This suggests that, by comparison with Equation \ref{mlum2}, \xco\ is
on average a factor of $1.7$ to $2.3$ larger than our adopted Galactic
value of $2\times10^{20}$ \xcounits\ (although with large error
bars). We will discuss the consequences for \xco\ and the presence of
any metallicity trends in the CO-to-H$_2$ conversion factor in
\S\ref{cotoh2}.

\subsection{Comparison with Complete Samples}

We outlined in \S\ref{introresol} some of the conclusions of the
recent study of the Local Group by \citet{BLITZ07}.  We employed here
the same datasets for M~31 and M~33, analyzing them with a later
version of the same CPROPS algorithm. While \citeauthor{BLITZ07}
emphasized completion to study the cloud statistics, the present
analysis made use of the available data with the best possible
combination of sensitivity and resolution, thus employing different
datasets for the Magellanic Clouds and IC~10. Furthermore, this study
incorporates a large number of galaxies in and beyond the Local Group
that were not analyzed in \citet{BLITZ07}. Many of the results,
however, are reassuringly similar: extragalactic GMCs broadly share
the same properties observed in Milky Way GMCs, although the clouds in
some galaxies (notably the SMC) appear to be somewhat different from
the remainder of the sample. We will argue later (\S\ref{smc}) that
this is probably caused by the low metallicity of the parent galaxy.

\section{Discussion}
\label{discussion}

\subsection{Implications for Photoionization-Regulated Star Formation}
\label{phototheory}

\begin{deluxetable*}{lcccccccccc}
\tabletypesize{\scriptsize}
\tablecolumns{11}
\tablewidth{0pt}
\tablecaption{\label{BIMATAB} Radio Observations}
\tablehead{
\colhead{Galaxy} & \colhead{\#} & \colhead{Trans.} & \colhead{R.A.} & \colhead{Declination} 
& \colhead{V$_{\rm LSR}$} & \colhead{$\log_{10}R$} & \colhead{$\log_{10}\sigma$} &
\colhead{$\log_{10}M_{lum}$} & \colhead{$\log_{10}M_{vir}$} &
\colhead{T$_B$} \\
\colhead{} & \colhead{} & \colhead{} & \colhead{(J2000)} & \colhead{(J2000)} 
& \colhead{(km s$^{-1}$)} & \colhead{(pc)} & \colhead{(km s$^{-1}$)} & \colhead{(M$_\odot$)} 
& \colhead{(M$_\odot$)} & \colhead{(K)}}
\startdata
\cutinhead{BIMA}
NGC185 & 1 & $1\rightarrow0$ & $ 00^{\rm h} 38^{\rm m} 56.4^{\rm s}$ & $ 48 \arcdeg 20 \arcmin 19.9 \arcsec $ & -194.1 &   \nodata  & $ 0.69 \pm  0.16$ & $  4.52 \pm   0.18$ &  \nodata  & $ 1.2$ \\
NGC205 & 1 & $1\rightarrow0$ & $ 00^{\rm h} 40^{\rm m} 24.0^{\rm s}$ & $ 41 \arcdeg 41 \arcmin 52.1 \arcsec $ & -248.9 &  $  1.43 \pm  0.09 $ & $ 0.78 \pm  0.11$ & $  5.02 \pm   0.12$ & $   6.01 \pm   0.24 $ & $ 1.0$ \\
NGC2976 & 1 & $1\rightarrow0$ & $ 09^{\rm h} 47^{\rm m} 23.6^{\rm s}$ & $ 67 \arcdeg 54 \arcmin 40.1 \arcsec $ &  -31.3 &  $  1.77 \pm  0.17 $ & $ 0.84 \pm  0.14$ & $  5.28 \pm   0.21$ & $   6.46 \pm   0.30 $ & $ 0.3$ \\
NGC2976 & 2 & $1\rightarrow0$ & $ 09^{\rm h} 47^{\rm m} 24.1^{\rm s}$ & $ 67 \arcdeg 54 \arcmin 33.9 \arcsec $ &  -24.8 &   \nodata  & $ 0.25 \pm  0.33$ & $  4.65 \pm   0.33$ &  \nodata  & $ 0.3$ \\
NGC2976 & 3 & $1\rightarrow0$ & $ 09^{\rm h} 47^{\rm m} 14.9^{\rm s}$ & $ 67 \arcdeg 54 \arcmin 44.0 \arcsec $ &  -19.5 &  $  1.75 \pm  0.18 $ & $ 0.52 \pm  0.14$ & $  5.43 \pm   0.11$ & $   5.82 \pm   0.30 $ & $ 0.4$ \\
NGC2976 & 4 & $1\rightarrow0$ & $ 09^{\rm h} 47^{\rm m} 18.4^{\rm s}$ & $ 67 \arcdeg 54 \arcmin 48.5 \arcsec $ &   -6.2 &  $  2.06 \pm  0.22 $ & $ 0.82 \pm  0.17$ & $  5.54 \pm   0.16$ & $   6.71 \pm   0.28 $ & $ 0.3$ \\
NGC2976 & 5 & $1\rightarrow0$ & $ 09^{\rm h} 47^{\rm m} 17.3^{\rm s}$ & $ 67 \arcdeg 54 \arcmin 59.2 \arcsec $ &    5.5 &  $  1.96 \pm  0.25 $ & $ 0.29 \pm  0.30$ & $  5.49 \pm   0.30$ & $   5.56 \pm   0.51 $ & $ 0.3$ \\
NGC2976 & 6 & $1\rightarrow0$ & $ 09^{\rm h} 47^{\rm m} 15.3^{\rm s}$ & $ 67 \arcdeg 55 \arcmin 04.9 \arcsec $ &   15.4 &  $  2.09 \pm  0.19 $ & $ 0.63 \pm  0.18$ & $  5.88 \pm   0.23$ & $   6.37 \pm   0.28 $ & $ 0.4$ \\
NGC2976 & 7 & $1\rightarrow0$ & $ 09^{\rm h} 47^{\rm m} 15.3^{\rm s}$ & $ 67 \arcdeg 55 \arcmin 16.7 \arcsec $ &   20.8 &   \nodata  & $ 0.68 \pm  0.29$ & $  5.51 \pm   0.36$ &  \nodata  & $ 0.3$ \\
NGC2976 & 8 & $1\rightarrow0$ & $ 09^{\rm h} 47^{\rm m} 13.0^{\rm s}$ & $ 67 \arcdeg 54 \arcmin 54.5 \arcsec $ &   -2.6 &   \nodata  & $ 0.63 \pm  0.13$ & $  5.18 \pm   0.20$ &  \nodata  & $ 0.3$ \\
\enddata
\tablecomments{Table \ref{BIMATAB} is published in its entirety in the electronic 
edition of the {\it Astrophysical Journal}.  A portion is shown here 
for guidance regarding its form and content.}
\end{deluxetable*}

\begin{deluxetable*}{lcccccccr@{$\pm$}l}
\tabletypesize{\scriptsize}
\tablecolumns{10}
\tablewidth{0pt}
\tablecaption{\label{restab} Galaxy-wide Measured GMC Properties}
\tablehead{
\colhead{Galaxy} & \colhead{Meas.} & \multicolumn{2}{c}{M$_{vir}/10^5$} & \multicolumn{2}{c}{$\log\left[\xco/2\times10^{20}\right]$}& \multicolumn{2}{c}{$\sigma_v R^{-0.5}$} & \multicolumn{2}{c}{$\Sigma_{\rm GMC}$} \\ 
& & \multicolumn{2}{c}{(M$_\odot$)} & \multicolumn{2}{c}{(\xcounits)} & \multicolumn{2}{c}{(km s$^{-1}$ pc$^{-0.5}$)} & \multicolumn{2}{c}{(M$_\odot$ pc$^{-2}$)}\\
& & \colhead{mean} & \colhead{range} & \colhead{mean\tablenotemark{1}} & \colhead{range\tablenotemark{2}} & \colhead{mean\tablenotemark{1}} & \colhead{range} & \multicolumn{2}{c}{}}
\startdata
IC~10    & 9 & $2.6$ & $0.2 - 5.6$  & $+0.28 \pm 0.07$ & $0.5 - 5.9$ &  $0.57 \pm 0.08$ & $0.34 - 0.94$ & 108 & 31   \\
NGC~205 & 1  & $10.4$ & \nodata     & $+0.99 \pm 0.27$ & \nodata     &  $1.16 \pm 0.25$ & \nodata       & 447 & 191  \\  
SMC     & 11 & $0.1$ & $0.02 - 1.2$ & $+0.43 \pm 0.03$ & $1.1 - 6.6$ &  $0.37 \pm 0.08$ & $0.18 - 0.75$ & 45 & 21    \\
LMC      & 7 & $1.3$ & $0.1 - 3.1$  & $+0.30 \pm 0.04$ & $1.1 - 3.0$ &  $0.61 \pm 0.10$ & $0.35 - 0.70$ & 123 & 39   \\
NGC~1569 & 1 & $2.1$ & \nodata      & $+0.52 \pm 0.78$ & \nodata     &  $0.72 \pm 0.44$ & \nodata       & 170 & 209 \\   
NGC~2976 & 7 & $26.4$ & $3.6 - 53$  & $+0.67 \pm 0.11$ & $1.2 - 15.1$&  $0.57 \pm 0.06$ & $0.20 - 0.90$ & 106 & 21   \\   
NGC~3077 & 4 & $16.0$ & $1.9 - 48$  & $+0.06 \pm 0.12$ & $0.7 - 1.5$ &  $0.71 \pm 0.08$ & $0.47 - 0.99$ & 166 & 39   \\   
NGC~4214 & 3 & $6.0$ & $4.3 - 6.9$  & $-0.07 \pm 0.17$ & $0.7 - 1.6$ &  $0.56 \pm 0.12$ & $0.48 - 0.94$ & 102 & 44   \\   
NGC~4449 & 1 & $85.3$ & \nodata     & $+0.08 \pm 0.19$ & \nodata     &  $0.56 \pm 0.09$ & \nodata       & 105 & 33  \\   
NGC~4605 & 3 & $24.2$ & $19 - 30$   & $+0.23 \pm 0.21$ & $1.2 - 2.2$ &  $0.37 \pm 0.08$ & $0.29 - 0.48$ & 44 & 18    \\   
M~31    & 44 & $5.3$ & $0.6 - 21$   & $+0.26 \pm 0.03$ & $0.5 - 5.0$ &  $0.72 \pm 0.03$ & $0.35 - 1.62$ & 172 & 15   \\
M~33    & 14 & $2.3$ & $0.9 - 8.0$  & $+0.12 \pm 0.06$ & $0.5 - 2.6$ &  $0.70 \pm 0.07$ & $0.46 - 2.40$ & 160 & 31   \\
\enddata
\tablenotetext{1}{Error-weighted mean and corresponding error.}
\tablenotetext{2}{Range for $\xco/2\times10^{20}$.}
\end{deluxetable*}

\citet{MCKEE89} discusses a self-regulated theory of star formation,
where photoionization caused by the interstellar radiation field
controls the rate at which clouds contract and form new stars. In this
theory clouds undergo contraction until the energy input by the newly
formed stars balances the gravitational collapse. Support against
contraction at the level of individual clumps is provided by their
magnetic field. This contraction happens at the ambipolar diffusion
rate, as the neutrals diffuse through the ions which are anchored by
the magnetic field. The rate of diffusion is determined by the
ionization fraction of the cloud, which in its outer regions (which
comprise most of its mass) is set by the ultraviolet interstellar
radiation field (ISRF), and in its inner regions is due to cosmic ray
ionization. The total rate of star formation of a cloud thus depends
critically on its extinction and radiation environment.  Clumps
immersed in weaker ISRFs require less extinction to collapse, and
vice versa.

These ideas have considerable implications for the structure of GMCs
as a function of environment, and it is of great interest to test
their predictions.  In the theory outlined by \citet{MCKEE89},
extinction is the key parameter that regulates the equilibrium of
clouds, not total column density. Accordingly, \citeauthor{MCKEE89}
proposed as one of the key tests of the theory that clouds in galaxies
with different metallicities and dust-to-gas ratios should obey
different {\em size-line width} relations, as

\begin{equation}
\sigma_v \approx 0.72\,\left(\frac{\bar A_V}{7.5 \delta_{gr}}\right)^{0.5}
R^{0.5} \ {\rm km~s^{-1}}.
\label{photosf}
\end{equation}

This expression is the analog of equation \ref{size_linewidth_eq},
where the numerical coefficient now incorporates the cloud mean
extinction, $\bar A_V$, and the ratio of \av\ extinction per hydrogen
nucleus relative to the Milky Way, $\delta_{gr}$, so that for
Galactic clouds we recover the standard {\em size-line width}
relation. The prediction from photoionization-regulated star formation
theory is that equilibrium molecular clouds in galaxies with low
values of $\delta_{gr}$ should have high velocity dispersions for a
given size. In other words, clouds in low metallicity environments
need to be more massive to attain the characteristic extinction in
their inner regions for a given ISRF.

\begin{figure*}
\plotone{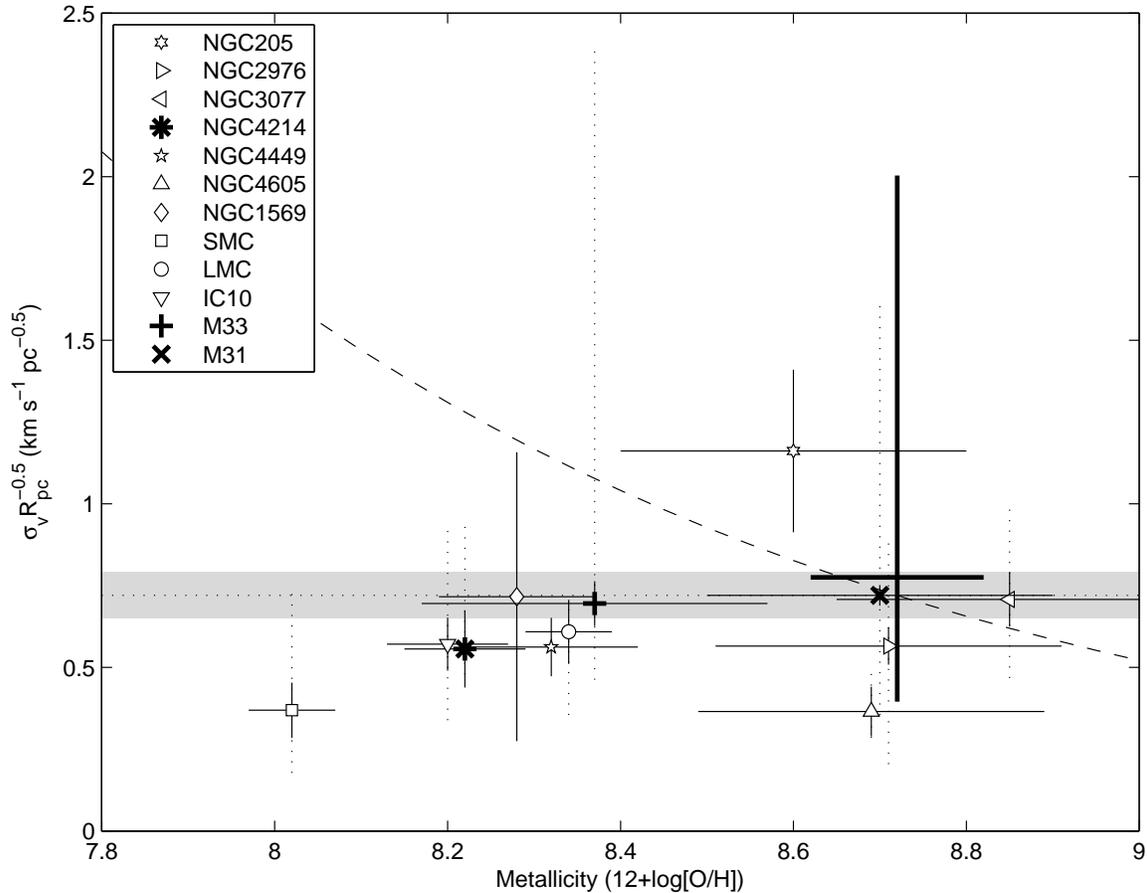}
\caption{Coefficient of the {\em size-line width} relation for the
different galaxies in our sample, as a function of metallicity.  Each
symbol represents the error-weighted average value of
$\sigma_v/R^{0.5}$ for all clouds in a galaxy. The horizontal bars
indicate the uncertainty in galaxy metallicity, while the vertical
fill bar correspond to the error in the error-weighted average and the
dotted vertical line illustrate the full range of values for
$\sigma_v/R^{0.5}$ for individual clouds within a galaxy. The thick
lines indicate the behavior for the Galactic sample of
\citet{SOLOMON87} corrected as described in the text, with the full
range shown.  The dashed line shows the expectation from equation
\protect\ref{photosf} assuming that the normalized extinction per
hydrogen nucleus ($\delta_{gr}$) is proportional to the source
metallicity, while the dotted horizontal line and gray area show the
value of canonical Galactic coefficient in Equation
\protect\ref{size_linewidth_eq}, $0.72\pm0.07$
km~s$^{-1}$~pc$^{-0.5}$. The fact that the dwarf galaxy points cluster
under the Galactic value is just another manifestation of their
tendency to fall under the Milky Way {\em size-line width} relation,
discussed in \S\protect\ref{sizelw}.\label{met_beta}}
\end{figure*}

Figure \ref{met_beta} shows the average coefficient of the {\em
size-line width} relation for each galaxy in our sample plotted as a
function of metallicity, which is a good proxy for $\delta_{gr}$ in
equation \ref{photosf}.  Neither the placement of individual GMCs in
the {\em size-line width} relation (Fig. \ref{size_linewidth}), nor
Fig. \ref{met_beta} show evidence of the systematic behavior predicted
by the theory. If anything, clouds in IC~10 ($\delta_{gr}\sim 0.3$)
and the SMC ($\delta_{gr}\lesssim0.2$), which should have
higher-than-Galactic velocity dispersions by factors of $\sim2-3$,
appear to have systematically lower-than-Galactic values of velocity
dispersion for a given size. Even if we discard the clouds that fall
under the Galactic {\em size-line width} relation as potentially out
of equilibrium or perhaps magnetically supported, the remaining clouds
do not show larger than Galactic velocity dispersion. It is important
to realize that several of these GMCs exist in environments of
larger-than-Galactic ISRF \citep[e.g.,][]{ISRAEL96}, so that the
observed lack of a trend in $\sigma_v R^{-0.5}$ is not directly caused
by a cancellation between the effects of ISRF and metallicity.

Early support for photoionization regulated star formation was found
by \citet{PAK98}, who used IRAS 60 and 100 $\mu$m data, a model of PDR
regions, and H$_2$ and [CII] measurements to derive the structure of
molecular clouds in the Magellanic Clouds. Their determination of
column densities, however, hinges critically on the estimate of the
radiation field, which is derived from IRAS measurements. In the
SMC they model dust temperatures in the range of $42-49$~K, higher
than the temperatures in the \citet{STANIMIROVIC00} map which employs
the same data. We use the IRIS data to check the temperature towards
LIRS36, one of their target regions, and find it to be 33~K instead of
the 45~K derived by \citet{PAK98}. Because the estimated radiation
field is very strongly dependent on the assumed dust temperature, this
alone suggests a $\sim 5$ times lower field than they adopt. This is
enough to bring their result into approximate agreement with our own,
and makes the radiation field in this region intermediate between
Galactic and 30 Doradus in the LMC. A further concern with this
analysis stems from the use of 60 $\mu$m data to establish a
temperature, as it is known that there is an important contribution
from stochastically heated small grains at this wavelength
\citep[particularly in the Magellanic Clouds,][]{BERNARD08}.  Studies
combining 100 $\mu$m data with $160-170$ $\mu$m
\citep{LEROY07,WILKE04} determine typical temperatures of $20-23$~K
for these SMC regions.

While our analysis does not directly support photoionization-regulated
star formation theory, there are several caveats that need to be
considered before rejecting it. First, strictly speaking the increased
column density prediction from photoionization-regulated star
formation theory applies to the structures that form stars. Thus, it
is possible that these observations do not sample the scales at which
the column density enhancements occur. Second, in addition to assuming
clouds in virial equilibrium, this prediction assumes that the ratio
of mean clump to cloud extinction is similar to Galactic clouds, and
that the characteristic density (at which cosmic ray ionization is
$\approx10^{-7}$) is a similar fraction of the cloud density as in
Galactic clouds.  Unfortunately, these parameters are almost
completely unconstrained for extragalactic GMCs. Third, metallicity
may not be a good linear proxy for $\delta_{gr}$. This parameter folds
in the effects of dust-to-gas ratio and extinction, and although
dust-to-gas ratio likely tracks metallicity the dependence of
extinction on metallicity may be important. In the SMC
$N(\hi)/\av\approx13.2\times10^{21}$ \percmsq~mag$^{-1}$, a factor of
6.5 larger than \citeauthor{MCKEE89}'s adopted Galactic value and
compatible with the ratio of metallicities between the SMC and the
Milky Way \citep{GORDON03}.  At ultraviolet wavelengths, however,
\citeauthor{GORDON03} find that $A_\lambda/\av$ is larger than
Galactic by factors of $2-3$. As a result of this, and assuming that
the ultraviolet extinction and not \av\ is the parameter that matters
at setting the cloud ionization level, we may probe a smaller range of
$\delta_{gr}$ than that assumed in Fig. \ref{met_beta}. Note that although
this consideration diminishes the discrepancy with the theoretical
predictions it does not eliminate it.  Fourth, equation \ref{photosf}
requires that CO is present throughout the molecular material so that
its velocity dispersion accurately samples the potential of the
GMC. This may not be the case in low metallicity environments if there
are large H$_2$ envelopes faint in CO surrounding the CO-bright
portions of the GMC, and it may constitute the biggest limitation in
our test of photoionization-regulated star formation. Nonetheless, the
result stands that there is no observed trend of increasing velocity
dispersion in the {\em size-line width} relation for decreasing
metallicity on the scales sampled --- {\em clouds show strikingly
similar properties in all galaxies}.

\subsection{The CO-to-H$_2$ Conversion Factor}
\label{cotoh2}

\subsubsection{Background}

Using $^{12}$CO observations to measure the amount of molecular
hydrogen requires assuming a CO-to-\htwo\ conversion factor
$\xco=N(\htwo)/\int I({\rm CO})\,dv$ \citep{LEBRUN83}. This factor
incorporates the effects of abundance, excitation, and cloud structure
averaged over a large area. The optically thick \co\ $1-0$ transition
can be used to trace the molecular mass of a cloud because it arises
from the surfaces of clump ensembles within the telescope beam: by
measuring \Ico\ we are counting clumps, and under very general
conditions the CO intensity will be proportional to the total
molecular mass of the clump ensemble. This is the essence of the CO
``mist'' model discussed by \citet{DICKMAN86}.  Because encompassing
all the relevant physics of the problem into ab initio calculations is
extremely difficult and not well constrained, astronomers rely on
empirical calibrations of \xco\ obtained using a variety of techniques
\citep[e.g.,][]{DICKMAN78,SANDERS84,BLOEMEN86,STRONG96,DAME01}.

It is reasonable to expect that \xco\ will depend on the local
properties of the interstellar medium (ISM); volume density,
temperature, radiation field, and metallicity
\citep*[e.g.,][]{KUTNER85,ELMEGREEN89,BELL06}. Observations throughout
the Galactic disk, however, constrain \xco\ within a narrow range
$\xco\approx(1.8\pm0.3)\times10^{20}$ \xcounits, with excursions of up
to a factor of $\sim2$ over this value, particularly at high latitudes
\citep{DAME01}.  This suggests that the local variations due to volume
density, temperature, and radiation field are unimportant when cloud
properties are averaged over a large area, which is the case for most
extragalactic observations. In cases where the average conditions are
very different from those in the Galaxy, however, differences in \xco\
become apparent. For example, mass estimates of molecular clouds
subjected to extreme radiation fields, such as those in starburst
environments, suggest that they have an \xco\ conversion factor 4 to
20 times smaller than in the Galaxy \citep[e.g.,][]{YAO03}. This can
be understood in terms of a larger emissivity per CO molecule in
starburst galaxies, stemming from a higher excitation state due to
larger physical temperatures and volume densities in these
environments \citep[e.g.,][]{WEISS01}. Another possibility is that a
key assumption underlying the CO-to-\htwo\ proportionality, that the
CO line width traces the \htwo\ mass, breaks down in very CO-rich
systems \citep*{DOWNES93}. Similarly, \xco\ conversion factors a few
times lower than that of the Galactic disk are commonly observed in
the centers of spiral galaxies, including our own
\citep*{SODROSKI95,ISRAEL06}.

A property of the ISM that is expected to have a dramatic impact on
the value of \xco\ is its heavy-element abundance
\citep{ISRAEL86,MALONEY88}. Unlike density or radiation field
fluctuations, metallicity does not average out over the large areas
sampled by a telescope beam. In fact, dwarf galaxies, which tend to
have subsolar metallicities, show little evidence for metallicity
gradients at least in the examples where such gradients could be well
measured
\citep[e.g.,][]{DUFOUR84,KOBULNICKY97,SKILLMAN03}. Metallicity affects
cloud structure in two ways: 1) directly, as smaller abundances of C
and O translate into lower CO yields, and 2) indirectly, as the
dust-to-gas ratio is lowered, which in turn diminishes the \htwo\
formation rates and the shielding of molecular gas from the
photodissociating effects of ultraviolet radiation.

The observational effects expected from these changes can be
summarized as follows: as metallicity decreases, decreasing dust
shielding pushes the C$^+$ to CO transition further into the molecular
gas. In terms of a spherical clump, the radius of the $\tau=1$ surface
for CO becomes smaller as the metallicity decreases.  Because at
high-enough extinction most of the carbon in the gas phase turns into
\co, which in turn becomes rapidly optically thick, the actual C/H
ratio of the gas has only a small effect on the position of $\tau=1$
surface. For a typical Galactic cloud, for example, the C$^+$/CO
transition occurs at $\av\sim2$ while the $\tau=1$ surface of CO
occurs at column densities $N_{\rm
CO}\sim8\times10^{15}$~\percmsq~(\kmpers)$^{-1}$ which translate into
visual extinction of $\av\sim0.015$ for a typical gas--phase carbon
abundance. Thus, almost immediately after becoming the dominant form
of carbon, \co\ turns optically thick.  Because of its large
abundance, \htwo\ is self--shielding in Milky Way molecular clouds
\citep{ABGRALL92,DRAINE96}. CO, however, is only mildly self--shielding,
although it also cross--shields with \htwo\ at column densities
$N(\htwo)\gtrsim 10^{21}$ \percmsq\ \citep{VANDISHOECK88}. If the
conditions are such that \htwo\ self-shielding dominates over dust
shielding at lower metallicities, the atomic to molecular \hi/\htwo\
transition will not move into the cloud for decreasing dust-to-gas
ratios. As a result, a molecular clump will not change its size with
decreasing metallicity but its \co\ emitting core will diminish,
yielding an increasing \xco\ conversion factor
\citep*[e.g.,][]{MALONEY88,BOLATTO99,ROLLIG06}.

\subsubsection{Existing Calibrations}
\label{calibrations}

There are several observational calibrations of \xco\ with
metallicity, $Z$, in the literature, showing a range of
behaviors. Most of the calibrations find an increasing \xco\ with
decreasing $Z$ although the rate of increase varies greatly depending
on the technique and the spatial resolution of the observations
used. 

A key datum in these calibrations is the measurement in the Small
Magellanic Cloud, as this is the galaxy with the lowest metallicity
where CO is reliably detected \citep[$Z_{\rm SMC}\approx0.2
Z_\odot$;][]{DUFOUR84,VERMEIJ02}.  In the first systematic study of
this galaxy, \citet{RUBIO93b} used single-dish CO observations and the
assumption of clouds in virial equilibrium to determine that
$\xco\sim60\times10^{20}$~\xcounits\ at a resolution of $\sim160$~pc,
while finding $\xco\sim9\times10^{20}$~\xcounits\ on scales of
$\sim15$~pc. \citet{MIZUNO01} revisited the estimate of \xco\ in the
SMC with new single--dish observations at a resolution of $\sim50$~pc
and found $\xco\sim10-50\times10^{20}$~\xcounits, in rough agreement
with the previous results. One important issue with these
determinations is that it is unclear whether the large scale
structures observed in CO are self-gravitating, virialized molecular
clouds. The applicability of the virial theorem on scales of 100~pc
and larger and the ability to accurately measure the size of the CO
structures, both necessary to apply equation \ref{virialmass}, are
problematic. Another issue with these determinations is that they do
not account for the finite angular resolution of the telescope, which
is in many cases comparable to the sizes of the clouds. This will
systematically bias these measurements in the direction of larger
virial masses and larger \xco.

Subsequently \citet{RUBIO04} used millimeter-wave continuum
observations to obtain the mass of the quiescent cloud SMCB1-1 and
compare it with its CO emission, reaching the conclusion that the
dust-derived mass is 7 to 20 times larger than the cloud virial
mass. They suggest that this \htwo\ is located in an envelope of the
cloud that does not emit brightly in CO. Very recently,
\citet{LEROY07} used new far-infrared (FIR) images of the SMC obtained
by the {\em Spitzer} Space Telescope to model the dust emission and
locally calibrate its dust-to-gas ratio, obtaining a map of \htwo\ in
this galaxy. Comparison with the CO emission indicates that the \htwo\
clouds are on average $\sim30\%$ more extended than the CO-emitting
regions, so that CO clouds are immersed in an extended molecular
envelope of \htwo. Furthermore, over the volume occupied by CO,
\citeauthor{LEROY07} find $\xco\sim60\times10^{20}$ \xcounits, while
the overall \xco\ for the entire SMC is approximately twice that.

Among the calibrations for \xco\ with metallicity, a number of them
rely on the assumption of virial equilibrium to obtain the molecular
mass of GMCs. \citet{WILSON95} used interferometric CO observations of
M~33 and dwarf galaxies in conjunction with the aforementioned
\citet{RUBIO93b} results to establish that $\xco\sim Z^{-0.67}$, where
$Z$ is measured using the oxygen abundance, as O/H. \citet{ARIMOTO96}
used several observations in the literature to determine that
$\xco\sim Z^{-1}$. Employing some of the same interferometric
observations analyzed here, \citet{WALTER01} determined a Galactic
value for \xco\ in NGC~4214, a galaxy where $Z\sim0.3\,Z_\odot$. A
result of 60\% of the Galactic value was obtained in NGC~3077
\citep{WALTER02}, a galaxy with approximately Galactic
metallicity. Similarly, \citet{ROSOLOWSKY03} used interferometric
observations of the entire disk of M~33 and found no dependence of
\xco\ on metallicity over a range of 0.8 dex in $Z$ (a factor of
6). All these studies are based on virial mass techniques.

Calibrations tend to be even more discrepant when other methods are
employed. Besides the aforementioned results by \citet{LEROY07},
\citet{ISRAEL97} used \hi\ and Infrared Astronomy Satellite (IRAS)
observations of several galaxies, estimating
$\xco\sim120\times10^{20}$~\xcounits\ in the SMC and finding a
metallicity dependence $\xco\sim Z^{-2.7}$, or $\xco\sim Z^{-3.5}$
when taking into account the local interstellar radiation field.
\citet{MADDEN97} used measurements of the FIR \cii\ \fscii\ transition
to estimate $\xco\sim50\times10^{20}$ in some regions of the low
metallicity dwarf galaxy IC~10, and over 100 times the Galactic value
overall ($Z_{\rm IC10}\sim0.3\, Z_\odot$).  \citet{IMARA07} used
stellar extinction in the LMC to measure
$\xco\sim9.3\times10^{20}$~\xcounits, toward the low end of virial
mass estimates for this source.  Finally, \citet*{BOSELLI02} found a
milder $\xco\sim Z^{-1}$ dependence using a combination of virial and
dust--continuum methods. In summary, there are large discrepancies
between the different authors and techniques, and in general estimates
based on FIR observations find stronger dependencies on metallicity
than those based on virial arguments, although typically they also
probe larger scales.

A problem with the available studies has been the lack of uniformity
in the datasets, methods, and analysis techniques. Unexpectedly, the
most uniform calibration in the literature \citep[that
of][]{ROSOLOWSKY03}, where all the clouds are at the same distance,
the spatial resolution is good (20 pc), and identical analysis is
applied to the data finds that \xco\ is independent of $Z$. Very
recent observations, however, cast doubts on the magnitude of the
metallicity gradient in this galaxy. \citet{ROSOLOWSKYSIMON07} find
that the metallicity gradient of M~33 is a factor of 3 shallower than
previously accepted, which implies that the Rosolowsky et al. (2003)
GMC data probes a considerably smaller range of $Z$ than previously
thought. Thus mild metallicity dependencies of \xco\ within this
galaxy may be masked by the internal scatter of the measurements.
Another potential issue is that similar gradients in the radiation
field and metallicity with galactocentric distance have opposite
effects and may conspire to produce an almost constant \xco\
\citep{ELMEGREEN89}.

\begin{figure*}
\plotone{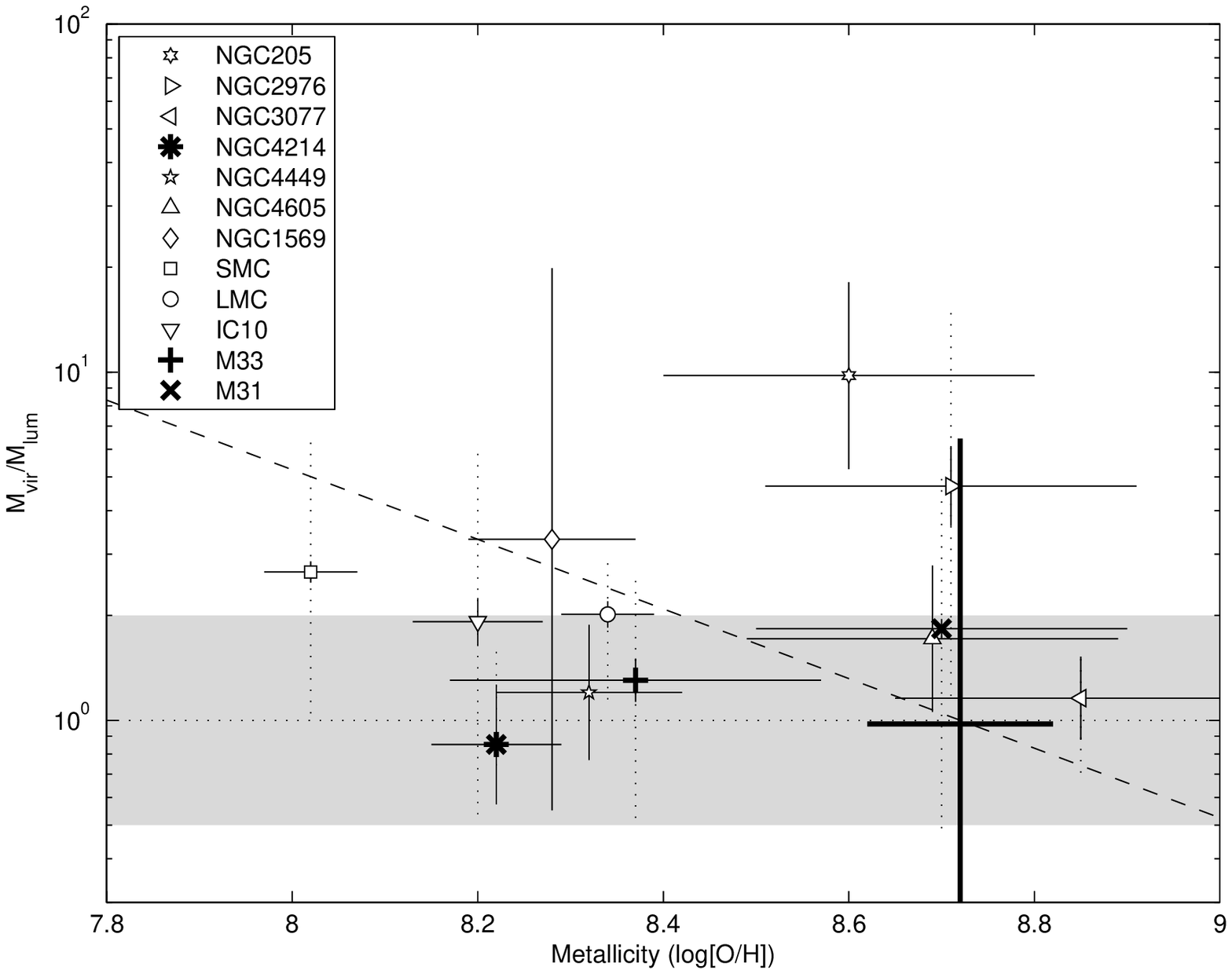}
\caption{Ratio of virial to luminous mass (i.e., \xco\ factor
including the contribution from He) as a function of metallicity for
the different galaxies in the sample.  The ratio is normalized to the
Galactic \xco, as all luminous masses are computed using our adopted
Milky Way value of the CO-to-\htwo\ conversion factor
($\xco=2\times10^{20}$ \xcounits, shown by the horizontal dotted
line).  The horizontal lines show the uncertainties in the metallicity
determination, while the vertical filled lines show the uncertainties
in the error-weighted \xco\ averages in each galaxy, and the dotted lines
show the range of ratios found for all clouds in each galaxy with more
than one identified GMC. The gray region indicates a factor of two
around Galactic \xco. The symbols represent the medians for all clouds
in each galaxy.  The thick lines show the results for the
\citet{SOLOMON87} sample of Galactic clouds with masses larger than
$10^{5}$ M$_\odot$. The dashed line illustrates $\xco(Z)\propto
Z^{-1}$. Note that the departure from the standard Galactic \xco\ for
the SMC is {\em in accordance with the Galactic relation between
luminosity and virial mass} (see Equation
\protect\ref{lco_mvir_eq}). Thus it is not directly associated with
metallicity, just a consequence of the smaller average mass of the SMC
clouds: similar mass clouds in the inner Milky Way show the same ratio
of virial to luminous mass.\label{metxco}}
\end{figure*}

\subsubsection{Our Results}
\label{ourxco}

The fact that the overwhelming majority of our extragalactic GMCs with
reliable property determinations are compatible with the empirical
Galactic relationship between $L_{\rm CO}$ and $M_{vir}$ described by
equation \ref{lco_mvir_eq} (\S\ref{luminosity_relations}) implies that
we do not observe extreme departures from a Galactic \xco.  Indeed,
Figure \ref{metxco} shows our results for the extragalactic
calibration of \xco\ with metallicity, using virial GMC masses.  The
gray region illustrates the approximate range of \xco\ in the Milky
Way found by \citet{DAME01}, a factor of 2 around the nominal
$2\times10^{20}$ \xcounits. The thick lines show the dispersion of
values for the GMCs in the \citet{SOLOMON87} sample with masses larger
than $10^5$ M$_\odot$. The symbols show the average of
$M_{vir}/M_{lum}$ for all GMCs in each galaxy in our sample, with the
corresponding vertical bars indicating the error and range of
$M_{vir}/M_{lum}$ within a galaxy.  The galaxy that shows the largest
departure from Galactic \xco\ is NGC~205, but with only one identified
cloud we lack significant statistics. The situation is similar for
NGC~1569. Note that although the GMCs in the SMC show an average \xco\
three times Galactic, they are not unusual compared with Galactic
clouds of the same mass (recall that Equation \ref{lco_mvir_eq} is not
a linear relation, and Galactic clouds with masses similar to our SMC
clouds have similar ratios of virial to luminous mass; c.f., Figure
\ref{lco_mvir}). Thus, GMCs in dwarf galaxies are remarkably
compatible with a Galactic \xco\ {\em independent of their
metallicity}, and any metallicity trends appear to be much weaker than
the dispersion of the measurements.

To further quantify this statement we have carried out a number of
fits to the data. We do not consider in the fits NGC~2976 (because of
its uncertain metallicity), and the Milky Way (which may suffer from a
systematic methodological offset). A least-squares bivariate fit
$\log\xco=a+b\log(O/H)$ with simultaneous errors in both axes to the
measurements for all remaining galaxies yields a slope
$b=-0.45\pm0.30$ and $\chi^2\approx21$, with errors estimated using
bootstrapping. Increasing all errors proportionally to obtain
$\chi^2\approx1$ yields $b=-0.23\pm0.25$. Similar fits to the data for
the galaxies with the largest number of GMC measurements (M~31, M~33,
IC~10, LMC, and SMC) yield $b=-0.46\pm0.30$ ($\chi^2\approx5$) and
$b=-0.24\pm0.26$ ($\chi^2\approx1$). These results, plus the
considerations in the paragraph above, confirm that there is no
measurable trend in the resolved \xco\ with metallicity present in the
data. As we discuss below, however, this does not imply that there is
no metallicity trend in the global ratio of CO-to-H$_2$ in galaxies.

What about the possibility of clouds being more massive than the
simple virial mass estimate? We already discussed the observation that
several clouds in IC~10 and the SMC are not quite compatible with the
Galactic {\em size-line width} relation, in the direction of too small
a line width for a given size. We will argue in \S\ref{smc} that a
possibility is that these clouds are transient structures not in
equilibrium that will collapse in a few Myr due to the lack of
turbulent support, or that they may be supported by magnetic fields in
larger proportion than the typical Galactic GMC.  In either case, such
clouds could be removed from the sample since their line width may
underestimate their true mass. Doing so slightly changes the averages
for the SMC and IC~10, but does not significantly alter the results.

Thus, a consistent and uniform analysis of the available data for
resolved GMCs shows no evidence for an increasing \xco\ with
decreasing metallicity. We note that this is unlike the scenario
mentioned before, where similar gradients in radiation field and
metallicity conspire to keep \xco\ approximately constant in the disks
of the Milky Way or M~33. Here we have galaxies (and particular GMC
complexes) that simultaneously have a higher-than-Galactic radiation
field and a lower-than-Galactic metallicity, yet a very similar \xco\
for their resolved GMCs \citep[e.g.,][]{ISRAEL96,MADDEN97,WILKE04}.

We emphasize that these results correspond to the resolved \xco.  As
we discussed in \S\ref{calibrations}, FIR observations show that in
galaxies such as the SMC, CO-bright cores are likely embedded in
extended envelopes of \htwo\ that do not emit in CO
\citep{MADDEN97,LEROY07}. The global molecular gas content of a galaxy
relative to its CO luminosity may well steeply scale with its
dust-to-gas ratio or metallicity, as it is strongly suggested by the
available data on a few of these objects
\citep{ISRAEL97,LEROY07}. Nevertheless, it appears that the structures
that we are able to identify as individual GMCs by means of their CO
emission have CO-to-\htwo\ conversion factors (as well as Larson
relations) that are approximately Galactic, independent of their
nebular metallicities.

We can understand the joint results from the CO and dust-continuum
observations in the following terms: in a low-metallicity environment
such as the SMC, CO-bright clouds are the innermost portions of
considerably larger H$_2$ structures mostly devoid of CO
emission. This scenario was suggested by \citet{RUBIO91} and
\citet{RUBIO93b} for the SMC and is consistent with calculations of
the effect of metallicity on the placement of the photodissociation
fronts \citep{MALONEY88,ELMEGREEN89}. It also has been suggested as
the situation in outer galaxy disks \citep*[but see also
\citealt{WOLFIRE08} for a recent determination of H$_2$ formation
rates in diffuse gas]{PAPADOPOULOS02}. It appears that \citep[as
described for Milky Way GMCs by ][]{HEYER04} these inner portions
approximately follow the Larson relations and, to CO observers capable
of resolving them, they appear very similar to Galactic GMCs. Thus an
approximately Galactic \xco\ factor correctly estimates the mass of
the CO-bright core, but underestimates considerably the mass of H$_2$
in the entire GMC. In this scenario the bulk of the mass has to be
located in an envelope, surrounding the CO emission. This requires
that the radiation field responsible for the photodissociation of CO
be external to the cloud, and that \htwo\ be effective at
self-shielding even in an environment with low dust-to-gas ratio and
consequently reduced \htwo\ formation rate.

\subsection{Brightness Temperatures}
\label{tpeak}

Column (11) in Table \ref{BIMATAB} gives brightness temperature at the
cloud peak, ${\rm T}_B$, for each GMC, a quantity set by the
excitation temperature, optical depth, and filling factor of
CO. Although not part of the Larson relations, this quantity is
accessible in extragalactic GMCs and may yield insight into the
physical state of clouds.

For Galactic clouds the average brightness temperature is $\sim 4$~K
\citep{SOLOMON87}. Many of our data exhibit notably lower ${\rm T}_B$
than this. This might indicate either lower excitation temperatures or
a different optical depth of CO, due to e.g. a lower cloudlet filling
factor or other geometry well below our resolution. However, the
simplest explanation is that we observe GMCs with a (spatially) large
beam and that beam dillution, i.e. a low CO filling factor within the
beam, lowers ${\rm T}_B$ in more distant systems.

\begin{figure*}
\plotone{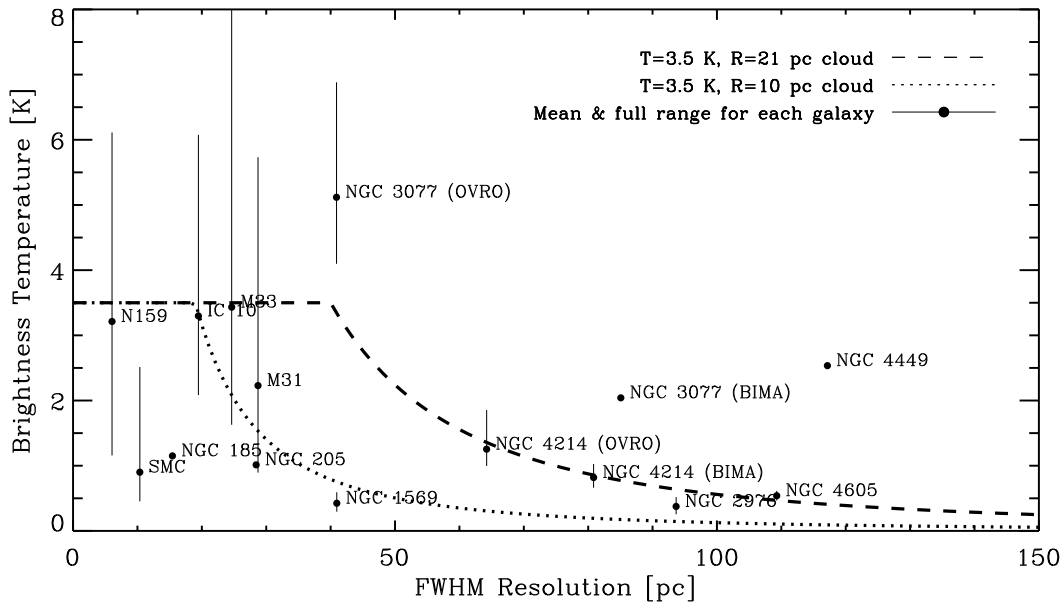}
\caption{Brightness temperatures in a GMC as a function of the spatial
  resolution of the data set. Circles show the mean value and solid
  bars the full range of brightness temperatures for each galaxy. The
  dashed line shows the expected curve for a uniform brightness
  ($T=3.5$~K) cloud with $R=21$~pc (a typical Galactic GMC). The
  dotted line illustrates the expected behavior for a $R=10$~pc
  cloud (a typical size for the SMC). Most galaxies are approximately
  compatible with the Galactic curve, but deviations exist: NGC~3077
  and NGC~4449 have brighter CO than other galaxies while the SMC,
  NGC~185, and NGC~205 exhibit low peak temperatures even at high
  spatial resolution. \label{fig_tpeak}}
\end{figure*}

Figure \ref{fig_tpeak} shows that beam dilution indeed explains many
of the the low observed ${\rm T}_B$. We plot the mean and full range
of ${\rm T}_B$ for each galaxy as a function of the spatial resolution
of the data. We also show the expected ${\rm T}_B$ for the case of a
uniform brightness cloud with FWHM size 40~pc (dashed line). The
dashed line is consistent with most of our data, including both Local
Group galaxies (the LMC, IC~10, M~33, and M~31) and more distant
dwarfs (NGC~2976, NGC~4214, NGC~4605). For the more distant dwarfs,
this highlights that while we resolve large ($\sim 100$~pc) structures
in the CO, these are likely blends of several GMCs.

A few extreme values of ${\rm T}_B$ cannot be immediately explained by
beam dilution. NGC~3077 appears unusually bright in CO when observed
with both BIMA and OVRO, perhaps a sign the that ongoing starburst at
the heart of this galaxy leads to higher excitation
temperatures. NGC~4449 shows a similarly high ${\rm T}_B$. In NGC~185,
NGC~205, NGC~1569, and the SMC (N83) ${\rm T}_B$ is low despite
excellent spatial resolution. In these galaxies, molecular clouds are
physically smaller, have lower area filling fractions of CO emission,
and/or have lower kinetic temperatures than in large galaxies.  The
first two might be caused by the low abundance of CO or by diminished
shielding from dust as discussed in previous sections. A lower kinetic
temperature remains a possibility, but is not supported by existing
studies at least in the Magellanic Clouds \citep*{BOLATTO05}. Clouds
in the SMC are indeed physically smaller than in most other galaxies,
but that effect does not completely account for the observed
differences (see dotted line in Figure \ref{fig_tpeak}). This
suggests that the CO emission has a lower area filling fraction: clouds
in some of these galaxies are porous structures, where CO arises only
from high $\av$ clumps.

\subsection{The departures in the Small Magellanic Cloud}
\label{smc}

We discussed in \S\ref{sizelw} the peculiar situation of several of
the least massive clouds in this study, mostly those belonging to the
SMC, that systematically exhibit velocity dispersions that are too
small for their sizes according to the Galactic {\em size-line width}
relation. We noted that, despite their small sizes, these objects
display exactly the opposite behavior from that observed in the outer
Galaxy, where small clouds are confined by the external pressure
\citep{HEYER01}.

Under the assumption of turbulence-supported clouds in virial
equilibrium following a $\sigma_v\propto R^{0.5}$ relation, the fact
that many SMC and a few IC~10 clouds lie a factor of two in $\sigma_v$
below the {\em size-line width} relation implies that their surface
densities are four times lower than Galactic clouds, yielding
$\Sigma_{\rm GMC}\sim45$ \msunperpcsq. Recall that, for Milky Way clouds, the
observed surface density translates into a visual extinction
$A_V\sim7.5$ through the cloud. In low-metallicity galaxies such as
the SMC and IC~10, where the dust-to-gas ratio is lower than in the
Milky Way by at least a factor similar to their heavy-element deficit,
a reduction by a factor of 4 in surface density is compounded by
another factor of $\sim4$ in the dust-to-gas ratio implying that we
would expect extinctions of order $A_V\sim0.5$ through the cloud, or
$A_V\sim0.2-0.3$ at the cloud center. We do not expect bright
$^{12}$CO emission at such low extinction, where most CO molecules
would be photodissociated. It is possible, however, that GMCs in the
SMC have a considerably lower CO area filling fraction than their
Galactic counterparts. Thus, although the average surface densities
and corresponding extinctions are are too low to allow the formation
of CO, it still exists in small, well shielded clumps within these
structures. We have discussed in the previous section that there is
some evidence along these lines, since clouds in the SMC indeed have
lower brightness temperatures suggestive of smaller CO area filling
fractions.

Alternatively, if the hypothesis of virialized clouds supported
chiefly by turbulent motions is not valid, the SMC and IC~10 GMCs that
fall under the {\em size-line width} relation show a deficit of
turbulent kinetic energy with respect to similar size Milky Way
clouds.  Since Galactic clouds are supported against collapse by a
combination of turbulence and magnetic fields with turbulence lending
the main support on the large scales \citep{MCKEE89,MCKEE07}, this
apparent deficit of turbulent energy would translate into rapid
collapse and subsequent star formation in a free-fall timescale
($t_{ff}\sim4.4\,(\bar n/100)^{-0.5}$ Myr, where $\bar n$ is the mean
volume density of the cloud; \citealt{MCKEE99}) unless there is
significant cloud support provided by a magnetic field that is
proportionally larger (or better coupled to the cloud material,
perhaps due to larger cloud ionization fractions) than in otherwise
similar Galactic structures.

To test this hypothesis it is necessary to obtain cloud mass
determinations independent of virial assumptions. This is usually
attained by employing the \xco\ factor, modeling molecular line
emission, or using measurements of the dust continuum and assuming a
dust-to-gas ratio and grain emissivity. The latter method was used by
\citet{BOT07}, who determined masses for several molecular clouds in
the SW region of the SMC. These authors found the dust-derived masses
for these clouds to be, on average, twice as large as their virial
masses. Under the assumption of long-lived clouds, they attributed
this fact to additional magnetic support by a $B\sim15$ $\mu$G
field. This field is similar to that present in smaller structures (a
few parsecs in size with masses $M<10^3$ M$_\odot$) in the Milky Way
\citep{BERTOLDI92}. Along similar lines, \citet{LEROY07} modeling of
the far-infrared emission in the SMC to derive molecular gas surface
densities found a mean surface density of molecular gas $\Sigma_{\rm
GMC}=180\pm30$ \msunperpcsq\ on 46 pc scales, very similar to that
observed in Milky Way GMCs. These observations suggest that surface
densities in the SMC are higher than what would be implied by cloud
line widths under the assumption of turbulence-supported virialized
clouds.

We argued in \S\ref{ourxco} that the joint CO and dust continuum data
in the SMC is best understood in terms of extended CO-faint \htwo\
envelopes surrounding CO-bright cores. In this interpretation, part
(possibly a large part) of the mass or surface density obtained by
dust continuum modeling is contributed by the extended \htwo\
envelope, and the mass of this envelope may not be accurately
reflected in the kinematics of the CO-bright core. Consequently, the
high dust-derived surface densities and low observed CO velocity
dispersions may not necessarily require an enhanced magnetic support.
We expect that further joint analysis of high spatial resolution FIR
and CO observations will shed light on the existence of extended
\htwo\ envelopes \citep{LEROY08}.

\section{Summary and Conclusions}
\label{summary}

We present and discuss a comprehensive set of high spatial
resolution observations of CO in low mass galaxies, obtained by a
combination of interferometer and single-dish instruments. Although
the data are heterogeneous, we analyze them in a consistent manner to
obtain GMC sizes, velocity dispersions, and luminosities.  The
analysis is performed using the algorithm described by
\citet{ROSOLOWSKY06}, which does a good job at removing the biases due
to dissimilar resolution and signal-to-noise.  We compare this uniform
dataset of resolved extragalactic molecular cloud properties against
those of GMCs in the three disk galaxies in the Local Group.  To do
so, we analyze the interferometric maps of M~31 \citep{ROSOLOWSKY07},
M~33 \citep{ROSOLOWSKY03}, and the sample of Galactic GMCs discussed
by \citet{SOLOMON87}.

The main result of this study is that, remarkably, {\em there are only
small differences between CO-bright GMCs in the Milky Way and GMCs in
galaxies with metallicities as low as $0.2Z_\odot$ subject to a
variety of physical conditions}. Our sample of extragalactic GMCs
follows approximately the same {\em size-line width}, {\em
luminosity-size}, and {\em luminosity-line width} relations as
Galactic clouds. Such uniformity may be in part responsible for the
observed invariance of the stellar Initial Mass Function in
galaxies. In any case, this result underscores that the Galactic
Larson relations provide a remarkably good description of CO-bright
Giant Molecular Clouds independent of their environment, at least in
the range of environments explored by this study.

Although the Larson relations are approximately Galactic there are
some significant departures. GMCs in dwarf galaxies tend to be
slightly larger than GMCs in the Milky Way, M~31, and M~33 for a given
CO luminosity or velocity dispersion. The largest departures occur in
the SMC, the galaxy with the lowest metallicity of the sample. A
possible interpretation, viable for most of our objects, is that GMCs
in small galaxies have on average a surface density of $\Sigma_{\rm
GMC}\sim85$ \msunperpcsq\ rather than the canonical $\Sigma_{\rm
GMC}=170$ M$\odot$~pc$^{-2}$ observed in the Galaxy. In the case of
the SMC, however, the implied surface densities, and consequently the
central extinction in the GMCs, would be too low to expect bright CO
emission. We explore three possibilities: these clouds are transient
structures not supported by turbulence and will collapse in a few Myr,
or else they are supported by magnetic fields in a larger proportion
than similar Galactic clouds, or maybe the kinematics of the CO cores
do not reflect the presence of massive \htwo\ envelopes.

Analysis of the properties of our sample of extragalactic GMCs shows
that they do not accord with simple predictions from the theory of
photoionization-regulated star formation \citep{MCKEE89}. In the
framework of this theory, clouds in lower metallicity environments
will have larger surface densities to attain similar extinction in
their central regions, which can then decouple from the magnetic field
and collapse. Such a trend would be evident in the coefficient of the
{\em size-line width} relation, as expressed by Equation
\ref{photosf}. As Fig. \ref{met_beta} illustrates, we see no evidence
for such a trend in our data. We point out four possible caveats with
these results: 1) it is possible that our observations do not probe
the spatial scales on which these enhancements may occur.  2) Perhaps
other parameters that enter in the theory (such as the cosmic ray
ionization rate, for example) change from galaxy to galaxy in a manner
that conspires to keep the coefficient of the {\em size-line width}
relation approximately constant. 3) Maybe our identification of
$\delta_{gr}$ with metallicity is incorrect and we sample a smaller
range of conditions than we think we do. Or, 4) the CO kinematics do
not trace the full potential of the cloud. In any case, we find a
$\sigma_v\,R^{-0.5}$ product that is approximately constant or even
decreasing for decreasing metallicity, contrary to expectations.

Finally, we address the matter of the dependency of the CO-to-\htwo\
conversion factor, \xco, on metallicity. We find that the
extragalactic GMCs in our analysis agree very well with the Galactic
{\em luminosity-virial mass} relation. Consequently, in our study of
resolved GMCs we find no measurable change of \xco: over a factor of 5
in metallicity most of our galaxies are compatible with a Galactic
$\xco=2\times10^{20}$ \xcounits\ within a factor of two, and there is
no discernible trend with metallicity. We emphasize, however, that
this measurement is relevant on the scales of the individual,
CO-bright GMCs. Studies in the FIR and at millimeter-waves in low
metallicity environments (particularly the SMC) suggest that these
clouds are embedded in larger \htwo\ envelopes that are not traced by
CO emission but contain an appreciable mass of molecular gas, at least
in the case of the SMC \citep{ISRAEL97,RUBIO04,LEROY07}.
These FIR results are approximately consistent with virial \xco\
estimates obtained on larger scales than the ones discussed in this
paper \citep{RUBIO93b,MIZUNO01,BOLATTO03}. We suggest that the FIR and
CO observations can be simultaneously understood in the following
terms: in low metallicity gas bright CO emission is relegated to the
density peaks. Observations that resolve those density peaks show
properties that are similar to those of Milky Way GMCs --- that is the
result of this study, and it is supported by studies of the structure
of Milky Way GMCs \citep{HEYER04,ROSOLOWSKY08}. On larger scales, the
cloud-to-cloud velocity dispersion as well as the dust opacity suggest
that there is a large mass of likely molecular material that is not
accounted for by the CO intensity. Such regions would be CO-bright in
objects of higher metallicity.

Studying the resolved properties of extragalactic GMCs is a
challenging undertaking with the current generation of instruments,
particularly for clouds in CO-faint dwarf galaxies. In the near future
the deployment of the ALMA interferometer will make it possible to
obtain more precise measurements on larger samples at farther
distances. Studies such as the one presented here suffer from an
important bias: we can only perform them on the brightest CO peaks.
The detection of clouds as faint as those in the SMC in CO, for
example, is currently impossible beyond the immediate neighborhood of
the Local Group. These improvements in instrumentation will soon
provide powerful tests of theories of star and molecular cloud
formation beyond our own Galaxy.

\acknowledgements We thank the referee, Christine Wilson, for her
thoughtful and constructive comments. We wish to thank Lisa Young,
Christopher L. Taylor, M\'onica Rubio, and Frank Israel for making the
data on NGC~185, NGC~205, NGC~1569, LIRS36, and LIRS49 available for
this study. We also want to thank Christopher McKee, Eve Ostriker,
Mark Wolfire, Mark Heyer, Mark Krumholz, M\'onica Rubio, Stuart Vogel,
and Andrew Harris for valuable discussions and comments on drafts of
this manuscript.  Alberto Bolatto wishes to acknowledge partial
support from the National Science Foundation grant AST-0540450.

\end{document}